\documentclass{aa}

\usepackage[utf8]{inputenc}
\usepackage{amsmath}
\usepackage{xcolor}
\usepackage{multirow}
\usepackage{xspace}
\usepackage{xstring}
\usepackage{supertabular}
\usepackage{tabularx} 
\usepackage{orcidlink} 
\usepackage{placeins}
\usepackage{hyperref}
\usepackage{enumitem}
\hypersetup{colorlinks=true, linkcolor=blue, filecolor=magenta, urlcolor=cyan, citecolor=blue}
\usepackage[varg]{txfonts}

\begin{document} 

\renewcommand{\thetable}{\arabic{table}}

   \title{Coronal electron density: Insights from radio and in situ observations and EUHFORIA modeling }
   
   \author{Ketaki Deshpande,\inst{1, 2}\orcidlink{0000-0001-6861-6328},
           Jasmina Magdalenic\inst{1, 2}\orcidlink{0000-0003-1169-3722},
           Immanuel Christopher Jebaraj\inst{3}\orcidlink{0000-0002-0606-7172}, 
           Senthamizh Pavai Valliappan\inst{1}\orcidlink{0000-0003-2413-7901}, 
           Antonio Niemela\inst{4,5}\orcidlink{0000-0002-3746-9246},
           Luciano Rodriguez\inst{1}\orcidlink{0000-0002-6097-374X},
           and Vratislav Krupar \inst{4, 5}\orcidlink{0000-0001-6185-3945}
           }
  
   \institute{Solar-Terrestrial Centre of Excellence – SIDC, Royal Observatory of Belgium, Avenue Circulaire 3, 1180 Uccle, Belgium,\\
    \email{ketaki.deshpande@oma.be}
         \and
              Centre for mathematical Plasma Astrophysics, Department of Mathematics, KU Leuven, Celestijnenlaan 200B, B-3001 Leuven, Belgium
         \and
              Department of Physics and Astronomy, University of Turku, 20500 Turku, Finland
         \and
              Goddard Planetary Heliophysics Institute, University of Maryland, Baltimore County, Baltimore, MD 21250, USA
         \and
              Heliospheric Physics Laboratory, Heliophysics Division, NASA Goddard Space Flight Center, Greenbelt, MD 20771, USA}

 \abstract 
   {The distribution of the coronal electron density at different distances from the Sun strongly influences the physical processes in the solar corona and it is therefore a very important topic in solar physics. The majority of the methods, including radio observations, used for estimation of coronal electron density were up to now not fully validated due to the absence of the in situ observations closer to the Sun. Consequently, space weather forecasting models which simulate coronal density lacked proper validation. Newly available Parker Solar Probe (PSP) in situ observations at distances close to the Sun, provide an opportunity to study the properties of plasma near the Sun and to compare observational and modeling results.}
  {The focus of this work is to study type III radio bursts, estimate their propagation path and validate coronal electron density obtained from radio, in situ observations and modeling with EUHFORIA (EUropean Heliospheric FORecasting Information Asset).} 
  {In this study of type III radio bursts observed during the second PSP perihelion, we employ radio triangulation and modeling to analyze coronal electron density. Using the radio triangulation method, we determine the 3D positions of the radio sources. Additionally, we utilize the state-of-the-art EUHFORIA model to estimate electron densities at various locations. The electron densities derived from radio observations and EUHFORIA modeling are then inter-validated with in situ measurements from PSP.}
   {We studied 11 Type III radio bursts during the second PSP perihelion, with radio triangulation showing their propagation path in the southward direction from the solar ecliptic plane. The obtained radio source sizes ranged from 0.5 to 40~$\deg$~(0.5~-~25~R$_\odot$), showing no clear frequency dependence. This indicates that scattering of radio waves was not very significant for the studied events and in this frequency range.
   
   A comparison of electron densities derived from radio triangulation, PSP in situ data, and EUHFORIA modeling showed a large range of obtained values. This result is influenced by the different propagation paths across different coronal structures, and model limitations. Despite these variations, EUHFORIA successfully identified high-density regions along type III burst paths, demonstrating its capability to capture large-scale density structures.}
  {Our study emphasize that type III bursts do not always follow the Parker spiral but instead trace distinct magnetic field lines which can be very differently oriented. The study shows rather constant radio source sizes and confirms that small-scale density fluctuations in PSP data remain relatively low. These two characteristics indicate that scattering effects do not significantly change observed radio source positions within the studied distances.}


\titlerunning{Coronal electron density: Insights from observations and modeling}
\authorrunning{K. Deshpande, et. al. }
\maketitle

\section{Introduction}
\label{sec:Intro}

\begingroup

\begin{table*}
\centering

\begin{tabular}{llllllllll}
\hline
 Event No. & Date  & \multicolumn{3}{c}{Observed Peak Time (UT)} & Associated  & \multicolumn{3}{c}{Complexity of Time Profile}  & PSP heliospheric    \\ 

 \cline{3-5}  \cline{7-9} 

    & April~2019 & Wind   & STA  & PSP & Phenomena
	&  Wind   & STA   & PSP  & distance (AU)        \\
 
    \hline
    \centering
 1  &  02   &   15:46      &  15:45       &   15:36       &  Jet    &   N    &  Y  & Y & 0.184171\\
 2  &  03   &   17:00      &  16:59       &   16:43       &  Jet    &   N    &  N  & Y & 0.171613\\
 3  &  03   &   19:00      &  18:59       &   18:43       &  Jet    &   N    &  Y  & Y & 0.170892\\
 4  &  03   &   22:34      &  22:33       &   22:26       &  Jet    &   N    &  N  & Y & 0.169674\\
 5  &  04   &   22:22      &  22:21       &   22:14       &  Jet    &   N    &  N  & Y & 0.165906\\
 6  &  05   &   01:24      &  01:23       &   01:22       &  Jet    &   Y    &  Y  & Y & 0.165952\\
 7  &  05   &   01:30      &  01:29       &   01:24       &  Jet    &   Y    &  Y  & Y & 0.165953\\
\textbf{8}  &  \textbf{05}   &   \textbf{17:06}      &  \textbf{17:05}       &   \textbf{16:58}       &  \textbf{Jet}    &   \textbf{Y}    &  \textbf{Y}  & \textbf{Y} & \textbf{0.168071}\\
\textbf{9}  &  \textbf{05}   &   \textbf{17:17}      &  \textbf{17:16}       &   \textbf{17:09}       &  \textbf{Jet}    &   \textbf{Y}    &  \textbf{Y}  & \textbf{Y} & \textbf{0.168113}\\
 10 &  05   &   20:08      &  20:07       &   20:01       &  Jet    &   Y    &  Y  & Y & 0.168842\\
 11 &  05   &   20:11      &  20:10       &   20:05       &  Jet    &   Y    &  Y  & Y & 0.168883\\
 \hline
\end{tabular}

\begin{tabbing}
\end{tabbing}
\vspace{-0.5 cm}
\caption {The characteristics of the studied type III radio bursts observed during the second PSP perihelion include: the observed peak time by Wind (708 kHz), STEREO-A (725 kHz) and PSP (721.875 kHz) spacecrafts, observed complexity in the time profile and information about the associated coronal phenomena.}
\label{table1}
\end{table*}
\endgroup

During the solar eruptive events such as Coronal Mass Ejections (CMEs) and solar flares we observe the wide spectrum of emission in different wavelengths including solar radio emission. As radio emission can provide information about background plasma processes and the eruptive phenomena, it is in the focus of numerous multi-wavelength studies of flare/CME events \citep[e.g.,][]{Pick98, Nindos08, Magdalenic14,Jebaraj20, Jebaraj23b, Kumari23}. Consequently, the study of radio emissions has a vital role in the field of solar physics for many decades \citep[see e.g.][]{Wild50a, Robinson94, Magdalenic14, Zucca18, Jebaraj20, Klein21}.

Solar radio bursts can be studied using four main types of observations: dynamic spectra, radio images, direction-finding and ionospheric scintillation data. Herein, we focus on the dynamic spectra and direction-finding observations. Dynamic radio spectra are frequency-time diagrams with color-coded intensity \citep[see,][]{Kundu61, VanHaarlem13, Marque18, Jebaraj23b}. Radio observations can be obtained using both ground-based radio telescopes and space-based receivers, each with their own limitations, but also offering valuable and complimentary insights into different types of bursts and their properties. Spectroscopic observations alone do not provide sufficient information to accurately determine the source location of the radio emission.
The ground based radio imaging observations provide us the radio source position in the 2D space, i.e. in the plane of the sky. Even more exact position of the radio sources can be obtained employing the direction-finding observations which provide information on the wave vectors indicating the direction of the radio emission in the three dimensional (3D) space \citep{Manning80, Reiner98}. 
Direction finding (DF) measurements from at least two spacecraft can then be used to triangulate the position of the source in 3D space \citep{Krupar12, Krupar14b, Krupar14a, Krupar15}, putting the radio emission in context with the flare/CME related processes \cite[see e.g.][]{MartinezOliveros12, Magdalenic14, Jebaraj20}.

Solar radio bursts are classified into five major groups (type I, II, III, IV and V), based on their spectral morphology as observed in the metric wavelength range \citep{Wild50a, Wild50b, Dulk70, Melrose80, NelsonM85}. This paper is centered on the study of type III solar radio bursts and estimation of the coronal plasma density at the source position of these radio bursts. Type III bursts are the most commonly observed type of solar radio emission. They are the electromagnetic signatures of near-relativistic electron beams \citep[$v_\mathrm{type III} \sim$ c/3,][]{SuzukiDulk85, Klassen03} that propagate along open or quasi-open magnetic field lines \citep{Pick06, Reid20}. Type III radio bursts are observed at fundamental and/or harmonic plasma frequency ($f_p$ \& $2f_p$), which involves generation and coupling of the Langmuir waves \citep{Ginzburg58, Melrose80, Robinson94, Voshchepynets15, Krasnoselskikh25}. As they propagate from high-to low-density regions in the solar corona they appear as a fast drifting lanes, from high to low frequency, in the dynamic spectra. Type III bursts can be observed all the way from GHz to kHz range \citep{Benz92, Melendez99, Leblanc95, Leblanc96, Dulk98, Krupar14b, Krupar14a}. 

Since the type III radio bursts are generated at the $f_p$ and/or $2f_p$ which is proportional to the electron density N$_e$, radio bursts can be used to estimate the coronal electron plasma densities at their source region, employing the radio images and DF data in metric and kilometric wavelength range, respectively \citep{Magdalenic10, Zimovets12, Magdalenic14, Jebaraj20, Jebaraj23b}. However, up to now in situ observations which would allow validation of the results obtained from radio observations were not available. The results derived from radio observations were therefore not well constrained. As a results of these uncertainties rather old 1D coronal electron density models \cite[e.g.,][]{Baumbach37, Newkirk61, Saito70, Leblanc98, Vrsnak04, Zucca18} are still most often employed.

Space weather forecasting models are a valuable tool for estimation of the coronal plasma characteristics, and in particular solar wind density and velocity. Such modeling results can be directly compared with different types of observations which allows, on one hand model validation and on the other hand it can support the observational results. Two of most frequently used state-of-the-art models of the background solar wind and CME propagation in the heliosphere are EUHFORIA \citep[European Heliospheric Forecasting Information Asset;][]{Pomoell18} and Enlil \citep{Odstrcil96, Odstrcil03, Odstrcil05}. In this work we will focus on the EUHFORIA model. The modeling accuracy of the solar wind plasma characteristics and arrival of high-speed streams and CMEs at Earth is often not very accurate, with the average forecasting errors as large as $\pm$~9~h \citep{Hinterr19, Samara21, Rodriguez24}.

The validation of the forecasting models is mostly done at 1~au where the in situ observations needed for model validation are available \citep[see e.g.][]{Owens08, MacNeice09b, Jian11, Reiss16, Reiss19, Hinterr19, Rodriguez24}. Only several works were devoted to the validation of the models at close to the Sun distances, \citep[such as e.g.,][]{McGregor11, Chen01, Leitner07}. Recent in situ observations by the Parker Solar Probe \citep[PSP;][]{Fox16} allowed model validation at various distances from the Sun \citep[][Senthamizh Pavai et al., in prep.]{Riley21, Wallace22, Samara24}. The novel PSP in situ data can also excellently serve for the validation of plasma parameters obtained from radio observations which also map the ambient solar wind characteristics at various distances from the Sun. We devote this study to the estimation and comparison of the coronal plasma density obtained through three different ways, modeling, in situ observations and radio observations.

Herein we present for the first time a comparison of coronal plasma densities estimated using type III radio bursts, radio triangulation method, in situ observations by PSP and modeling with EUHFORIA. This approach allows us to address uncertainties of both the radio triangulation technique and the EUHFORIA modeling together. The paper is structured as follows: The data used in this study is described in Section~\ref{sec:Data}. We discuss the method of radio triangulation and space weather model EUHFORIA in Section~\ref{sec:Method}, while Section~\ref{sec:density} elaborates on the different method of obtaining electron densities. Section~\ref{sec:case} presents the analysis of the selected events. Section~\ref{sec:triangresults} discuss the triangulation results, while Section~\ref{sec:comparison} addresses the results related to comparison of densities obtained by different methods and it summarizes the uncertainties induced by each method. The discussion and conclusions are presented in Section~\ref{sec:conclusion}.

\section{Data description}
\label{sec:Data}

In order to give the context on the coronal eruptive processes associated with the studied radio bursts we employed multi-wavelength analysis. Together with the radio observations we employed white light, Extreme Ultraviolet (EUV)  and Soft X-ray (SXR) observations. The main input for modeling with EUHFORIA were synoptic magnetograms, while in situ observations were used for the validation of modeling results and comparison with the coronal electron density obtained from radio observations. 

\subsubsection*{EUV and SXR observations:} The low solar corona was analyzed using Extreme Ultraviolet (EUV) observations from two instruments: \\
a) EUVI \citep[Extreme Ultra Violet Imager;][]{Howard08} on-board STEREO-A \citep[Solar TErrestrial RElations Observatory Ahead;][]{Kaiser08} provided images in four channels: 171, 195, 284, and 304~\AA. \\
b) AIA \citep[Atmospheric Imaging Assembly;][]{Lemen12} onboard SDO \citep[Solar Dynamics Observatory;][]{Pesnell12} provided observations in 7 different channels: 94, 131, 171, 193, 211, 304, and 335 \AA.\\
We also employed  soft X-ray ray (SXR) Observations obtained from the GOES \citep[Geostationary Operational Environmental Satellite;][]{Garcia94} to identify the flaring activity possibly associated with studied radio bursts.

\subsubsection*{White light observations:}
In order to study dynamics of the higher corona, such as possibly associated CMEs, we used coronagraphic observations from the STEREO-A and also LASCO \citep[Large Angle Spectrometric COronagraph;][]{Brueckner95} on board SOHO \citep[Solar and Heliopsheric Observatory;][]{Domingo95}. SOHO/LASCO includes externally occulted coronagraphs C2 (\mbox{1.5\,--\,6}~$R_\odot$) and C3 (\mbox{3.7\,--\,30}~$R_\odot$) while STEREO-A provides two field of view, COR1 (\mbox{1.5\,--\,4}~$R_\odot$) and COR2  (up to 15~$R_\odot$). 

\subsubsection*{Radio observations:}
The main focus of this paper are type III radio bursts observed by space-based radio instruments on board Wind \citep{Bougeret95}, STEREO-A and PSP \citep[Parker Solar Probe;][]{Fox16}. We have used two different types of observations: dynamic radio spectra and DF data. Wind/Waves instruments provide dynamic radio spectra obtained from two receivers RAD1 and RAD2 covering the frequency ranges of \mbox{20\,--\,1040\,kHz} and \mbox{1075\,--\,13825\,kHz}, respectively. DF measurements are available in discrete frequencies between \mbox{100\,--\,1040\,kHz} with temporal resolution of 45~s. This study uses the publicly released Wind/WAVES RAD1 direction-finding dataset, Version 01 \citep[released 2022;][]{Bonnin2022}. 
Dynamic spectra by two receivers LFR and HFR (Low \& High Frequency Receiver) onboard STEREO/Waves cover together the frequency range of \mbox{40\,--\,16000\,kHz}. Similarly, DF measurements made by STEREO/Waves are available in a somewhat smaller frequency range of \mbox{125\,--\,1975\,kHz} and with larger time resolution (30~s) at number of discrete frequencies \citep{Krupar22}. \\
The Radio Frequency Spectrometer \citep[RFS;][]{Pulupa17} part of the FIELDS \citep{Bale16} instrument suite onboard PSP provided spectroscopic observations in the range of \mbox{10.5\,--\,19000.2\,kHz} by combining both the High/Low Frequency Receivers (HFR/LFR). The full averaged spectra is obtained at a 7~s cadence during PSP's close encounters. 

\subsubsection*{Magnetogram:}

We used the synoptic magnetogram maps provided by NSO/GONG \citep[National Solar Observatory's Global Oscillation Network Group;][]{Harvey96} and ADAPT \citep[Air Force Data Assimilative Photospheric Flux Transport;][]{Hickmann15} maps as a main data input to the EUHFORIA model. These two type of magnetic maps will be further referred as GONG and ADAPT respectively. 

\subsubsection*{In situ data:}

The quasi-thermal noise (QTN) spectroscopy allows us to obtain the electron density through an accurate electrodynamic proxy by making use of the measurement of the plasma frequency by LFR/RFS \citep{Moncuquet20}. The RFS receiver has the optimal conditions for accurate measurement during the perihelion passages \citep{Pulupa17} which makes ``PSP density'' values used in this work quite reliable. PSP also has another instrument onboard which measures the plasma density, SWEAP \citep[Solar Wind Electrons Alphas and Protons;][]{Kasper2016}. SWEAP’s Faraday cups and electrostatic analyzers may underestimate electron density due to inability to fully capture strahl and halo electrons, especially in regions with low angular coverage; and the use of assumptions on the distribution function shape (often Maxwellian), which may not hold in turbulent or anisotropic conditions. In contrast, QTN spectroscopy uses thermal noise spectra, which are directly proportional to the electron plasma frequency, and thus offer absolute density estimates, especially in regimes where the plasma is not fully sampled by SWEAP. As this is also supported by number of studies  \citep[e.g.,][]{Meyer01,  Moncuquet20, Liu21, Martinovic22} we choose to employ the data from FIELDS instrument which is dedicated for the study of electrons, the prime focus of this paper.

\section{Methodology}
\label{sec:Method}

\subsection{Radio triangulation}
\label{sec:triangulation}

Mapping radio source positions in interplanetary space can be done employing different methods \citep[see e.g.,][]{Reiner09, Alcock18, Zhang19, Musset21, Badman22, Canizares24}. The majority of these methods while estimating the position of the radio sources and the direction of the propagation of the radio emission consider some approximations and simplifications which reduce the 3D methods to the 2D space. TDOA (time-difference-of-arrival) method by \cite{Badman22} assumes the location of the radio sources in the ecliptic plane along with the observing spacecraft. Quite similar to TDOA is BELLA (BayEsian LocaLisation Algorithm) method \citep{Canizares24} which is also reduced to the 2D space. The methods by \cite{Reiner09} and \cite{Musset21} bring rather large uncertainty in the source position estimation assuming the coronal electron density in order to obtain the radial distance of the sources from the Sun. Another method which makes assumption on the coronal electron density model is SEMP (Solar radio burst Electron Motion Tracker) by \cite{Zhang19}. This method also assumes the propagation of the radio sources along the Parker spiral with the constant speed. We note that accurately obtaining not only the propagation direction but also the 3D radio source positions is possible exclusively employing, so called, radio triangulation method \citep[e.g.,][]{Reiner98, Krupar14b, Krupar14a, Magdalenic14, Jebaraj20}. We consider this method for obtaining source position in 3D space, as it is the most reliable one since it does not include additional assumptions like most of the other methods.
 
In this study, we will further use 3D radio source positions to map the coronal electron density along the type III radio bursts propagation path. 

In order to employ radio triangulation and estimate the radio source positions in the 3D space we need goniopolarimetric (GP) or direction-finding observations from at least two or more widely positioned spacecraft. Using the DF methods on the GP observations, i.e., the flux, azimuth and colatitude, we can estimate the direction of arrival of an incident electromagnetic wave. Due to differences between STEREO and Wind spacecraft different methods need to be applied to the DF observations in order to obtain the information about the wave vectors. The spinning demodulation method \citep{Gurnett78, Manning80} was then used for the Wind data and singular value decomposition (SVD) \citep{Cecconi05, Cecconi08, Krupar12} for the STEREO data to obtain the wave vectors from the two different view points. Although the GP observations start around 1975~kHz, in this study we considered Wind \& STEREO-A frequency pairs at 425/428, 525/548, 575/548, 625/624, 675/624, 725/708, 775/708~kHz, respectively. 

\begin{figure}[ht!]
    \centering
    \includegraphics[width=9cm]{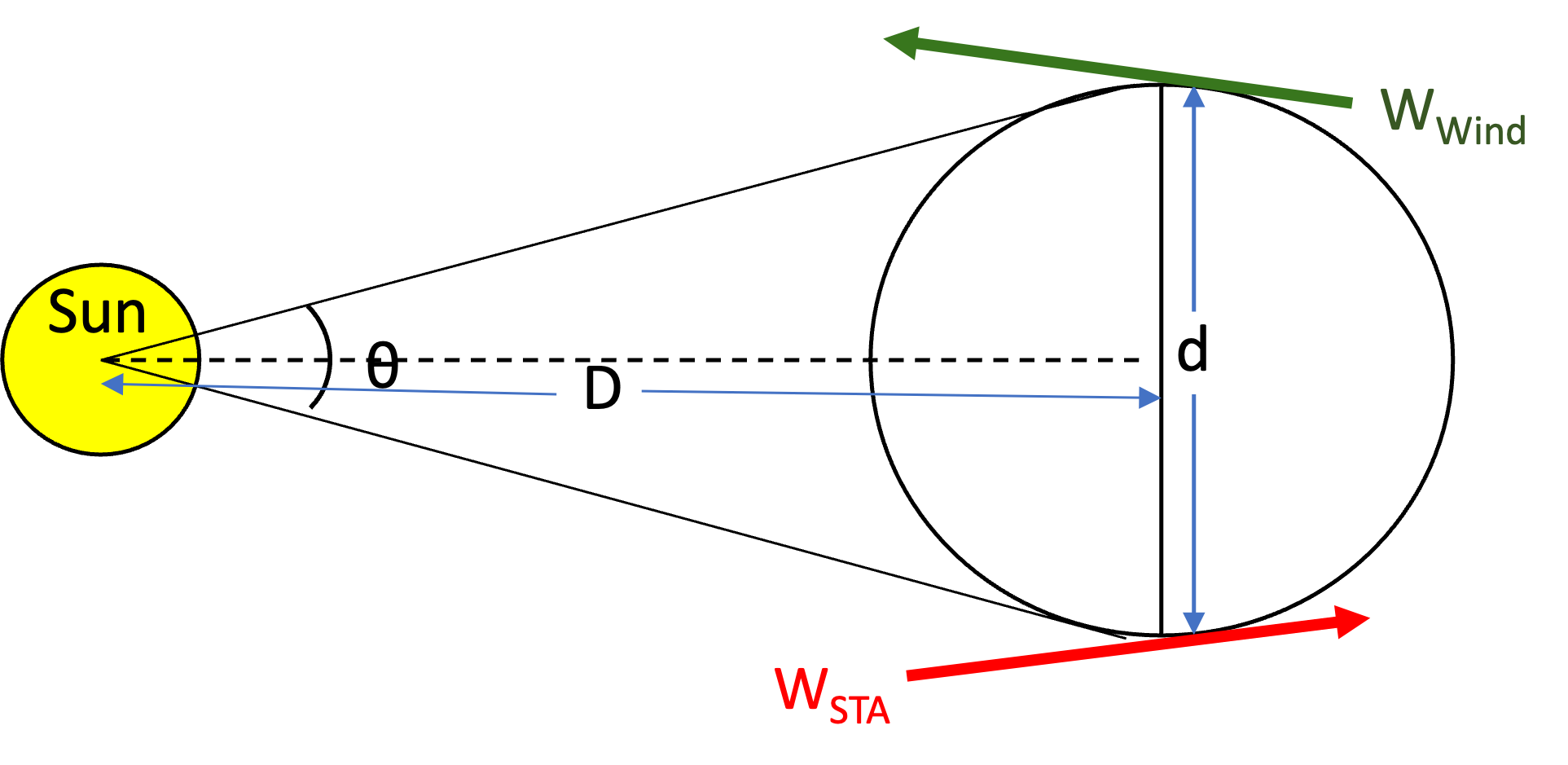}
     \caption{Schematic illustration of the radio source size, calculated in degrees from the shortest distance between the wave vectors $W_{wind}$ and $W_{STA}$. The distance d is derived from the Wind and STEREO-A DF observations. The sizes of the Sun, radio source and the distance between them are not in the real proportions.}
              \label{Fig: illustration}
    \end{figure}

The frequency range of roughly \mbox{400\,--\,800}~kHz was selected in order to assure that the fundamental emission is a dominant radio emission component with also often smaller and better defined source sizes as discussed by \cite{Dulk84, Melrose80}. The novel PSP data provide more observational evidence on the fundamental and harmonic of the type III radio bursts appearance. \cite{Jebaraj23b} shows that large fraction of type III bursts seen by PSP consists of fundamental and harmonic component. However, due to the proximity of the local plasma frequency,  the type III bursts below 1~MHz are not clearly distinguishable in the dynamic spectrum \citep[see e.g. Fig.~1 in][]{Jebaraj23b}. On the other hand, type III bursts examples in the study by \cite{Chen24} show rather clear dominance of the fundamental emission band below 1~MHz. Therefore, considering that majority of the type III burst below conservative value of 800~kHz, even smaller than before mentioned 1~MHz, is a reasonable assumption. This criteria is therefore also favorable for obtaining the electron density in the 3D space using radio source positions as accurate as possible. Once we obtained the wave vectors for the selected frequency pairs, we consider that the radio source position is within the shortest distance between these two wave vectors. We note that the radio source size could be actually smaller than the distance between these two wave vectors. Similarly to the previous studies \citep{Magdalenic14, Jebaraj20} the radio emission apparent source region is then defined as a sphere sketched in the (Fig.~\ref{Fig: illustration}). We estimate this apparent source size in degrees as, the angle $\theta$ projected on the solar surface by the shortest distance between the wave vectors d, expressed as,

  \begin{equation}  \label{eq:1}
      \theta = 2\mathrm{tan}^{-1}\left( \frac{d/2}{D}\right)
  \end{equation}

Where, D is the distance from the Sun to the radio source position.
We note that the spherical source shape approximation, used here also to estimate the source size and volume (Fig.~\ref{Fig: distances}b and \ref{Cell_size}) is  commonly used in the literature as a first-order approximation for the radio source shape \citep{Reiner98, Saint-Hilaire13, Magdalenic14, Jebaraj20}. 
This source shape approximation implies also the homogeneous intensity over the source region.
 
\begin{figure*}
    \centering
    \includegraphics[width=18cm]{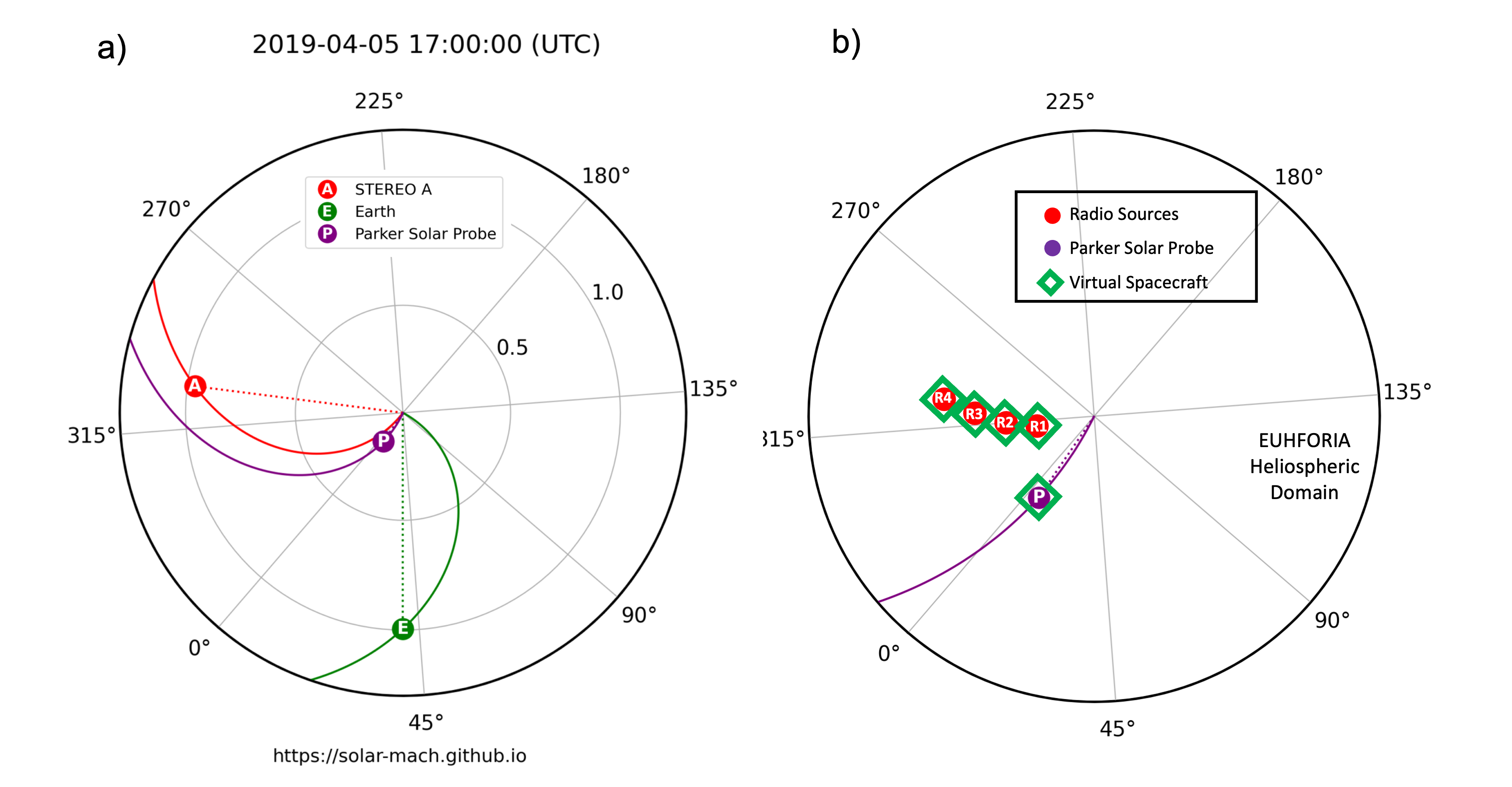}
     \caption{Panel (a) shows spacecraft locations in equatorial plane during the type III burst on 05~April~2019 ($\sim$ 17:00 UT). Panel (b) sketches the part of EUHFORIA's modeling domain up to 0.5~au. It also shows the arbitrary virtual spacecraft positions at which we consider modeled time series of the solar wind parameters. }
              \label{Fig: position}
    \end{figure*}

\subsection{EUHFORIA}
\label{sec:Euhforia}

One of the recently developed and dynamically evolving model of the solar wind and CMEs is a 3D magntohydrodynamic  (MHD) model EUHFORIA \citep{Pomoell18}. This data-driven forecasting model computes the time evolution of the inner heliospheric plasma. The ideal MHD equations are used to build the model that self-consistently simulates the solar coronal plasma. The main input to the model are synoptic magnetograms and the output is in the form of time series and 3D simulation results. EUHFORIA consists of two different domains: coronal and inner heliospheric domain, and a CME insertion part.  

The coronal model in EUHFORIA provides the boundary conditions at 0.1~au, so called inner boundary of EUHFORIA, necessary for modeling in the inner heliospheric domain. EUHFORIA's coronal model is based on the potential-field source-surface extrapolation model \citep[PFSS;][]{Altschuler69, Schatten69} followed by the Schatten Current Sheet model \citep[SCS;][]{Schatten71} and Wang-Sheeley-Arge model \citep[WSA;][]{Wang90, Arge00}. The heloispheric model reconstructs the evolution and propagation of the coronal plasma covering the distances from 0.1 to 2~au. Different CME models can be introduced in the heliospheric domain of EUHFORIA to study the CMEs propagation and forecast their arrival at Earth \citep{Maharana22, Niemela23, Rodriguez24, Valentino24}. 

While first accuracy testing  of the solar wind modeling was done at 1~au \citep[e.g.][]{Hinterr19, Asvestari19, Asvestari20}, recently available PSP in situ observations opened the possibilities for model validation at various distances close to the Sun  \citep{Samara24}. Validating and improving the solar wind modeling is important since it strongly affects also the propagation of CMEs and their interaction with the solar wind (Valentino et al., in prep.).

\section{Type III bursts and coronal electron density}
\label{sec:density}

During PSP's second close encounter, when the probe reached distances of up to ${\sim}0.16$~au from the solar surface, it observed several groups of type III radio bursts. These bursts were also detected by STEREO-A and Wind at ${\sim}1$~au. We inspected the entire three week period of the perihelion (\mbox{25~March\,--\,14~April,~2019}) in order to isolate type III radio bursts suitable for our study. This period is ideal given the near-quadrature alignment between Wind and STEREO-A spacecraft, 97$^\circ$ (Fig.~\ref{Fig: position}), which is best suited for the radio triangulation method \citep{Reiner09, Magdalenic14}.

 \begin{figure*}[ht!]
    \centering
    \includegraphics[width=17 cm]{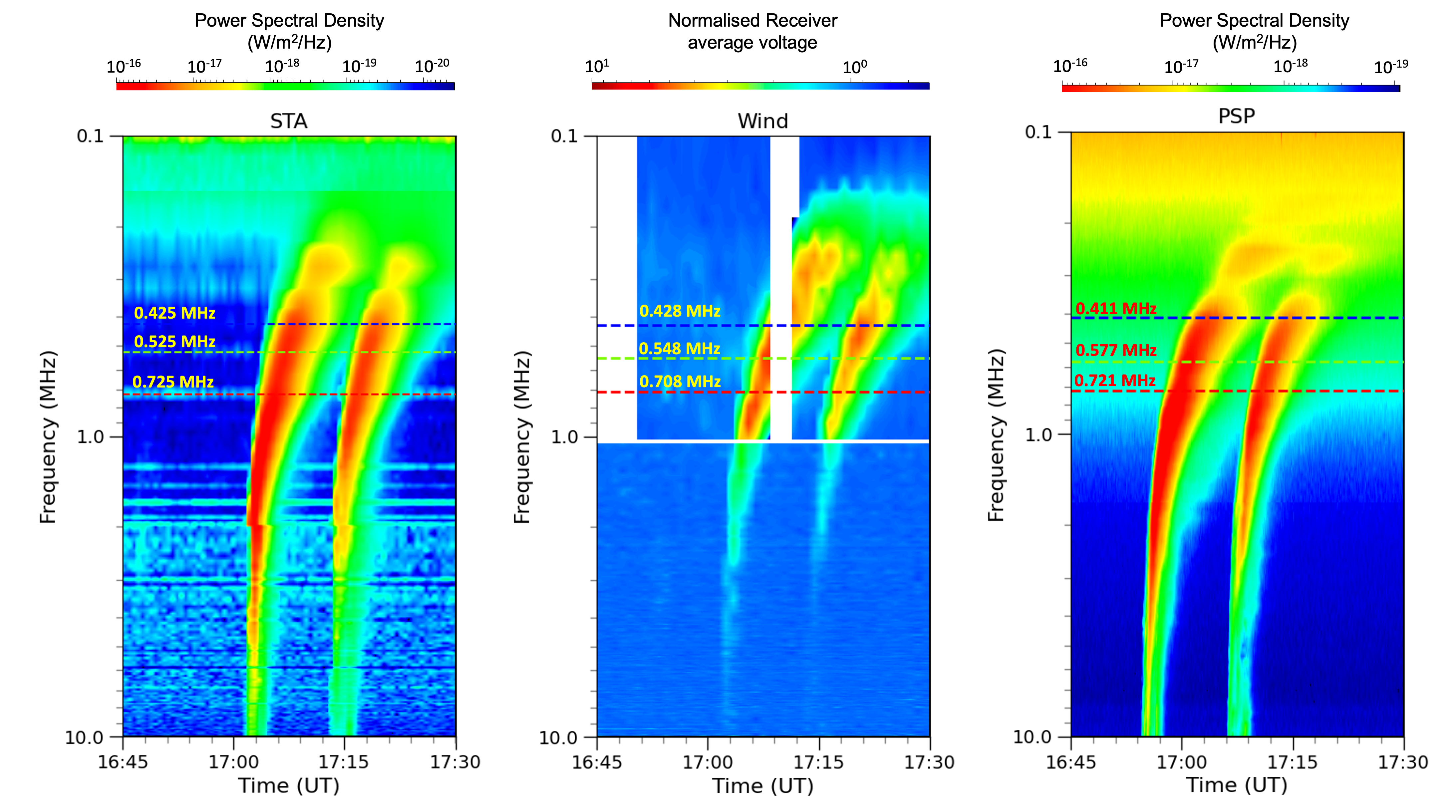}
     \caption{Dynamic radio spectra of the events 8 and 9 observed on 05~April~2019 at $\sim$ 17:00~UT. Events 8 and 9 (see also \autoref{table1}) were observed by Wind, STEREO-A, and Parker Solar Probe, from left to right panel. The dashed lines in red, green, and blue indicate the direction-finding frequencies for the Wind and STEREO-A spacecraft. For comparison similar frequencies are shown for PSP observations. }
              \label{Fig: Radio Spectra}
    \end{figure*}

 \begin{figure*}
    \centering
    \includegraphics[width=18cm]{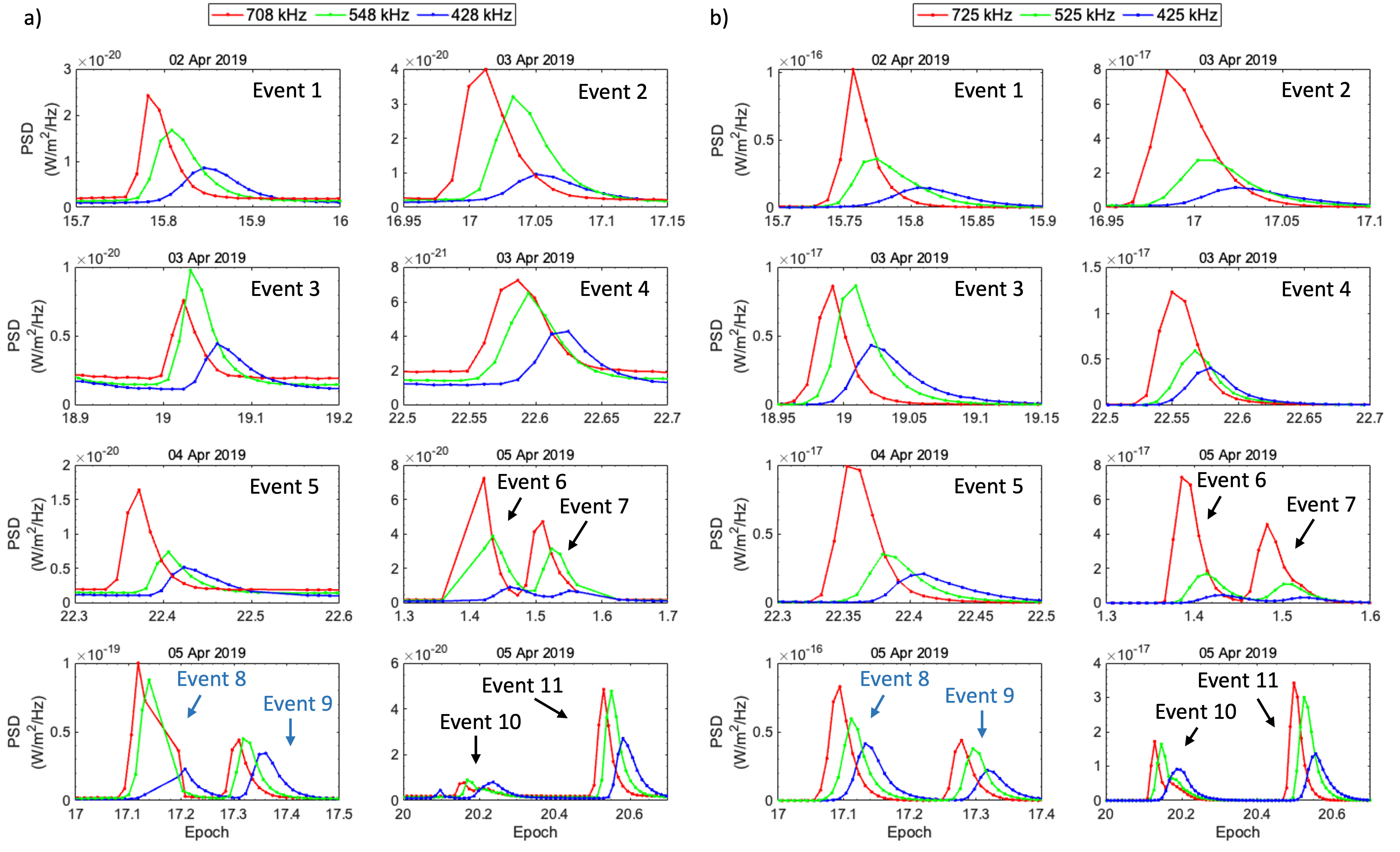}
    \caption{Panel (a) shows time profiles at 708~kHz (highest), 548~kHz (middle) and 428~kHz (lowest) direction-finding frequencies for all the studies type III radio bursts observed by Wind spacecraft. Panel (b) presents time profiles at 725~kHz (highest), 525~kHz (middle) and 425~kHz (lowest) direction-finding frequencies for all the studied type III radio bursts observed by STEREO-A spacecraft.}
              \label{Fig: Time Profiles}
    \end{figure*}

\begin{figure*}
    \centering
    \includegraphics[width=15cm]{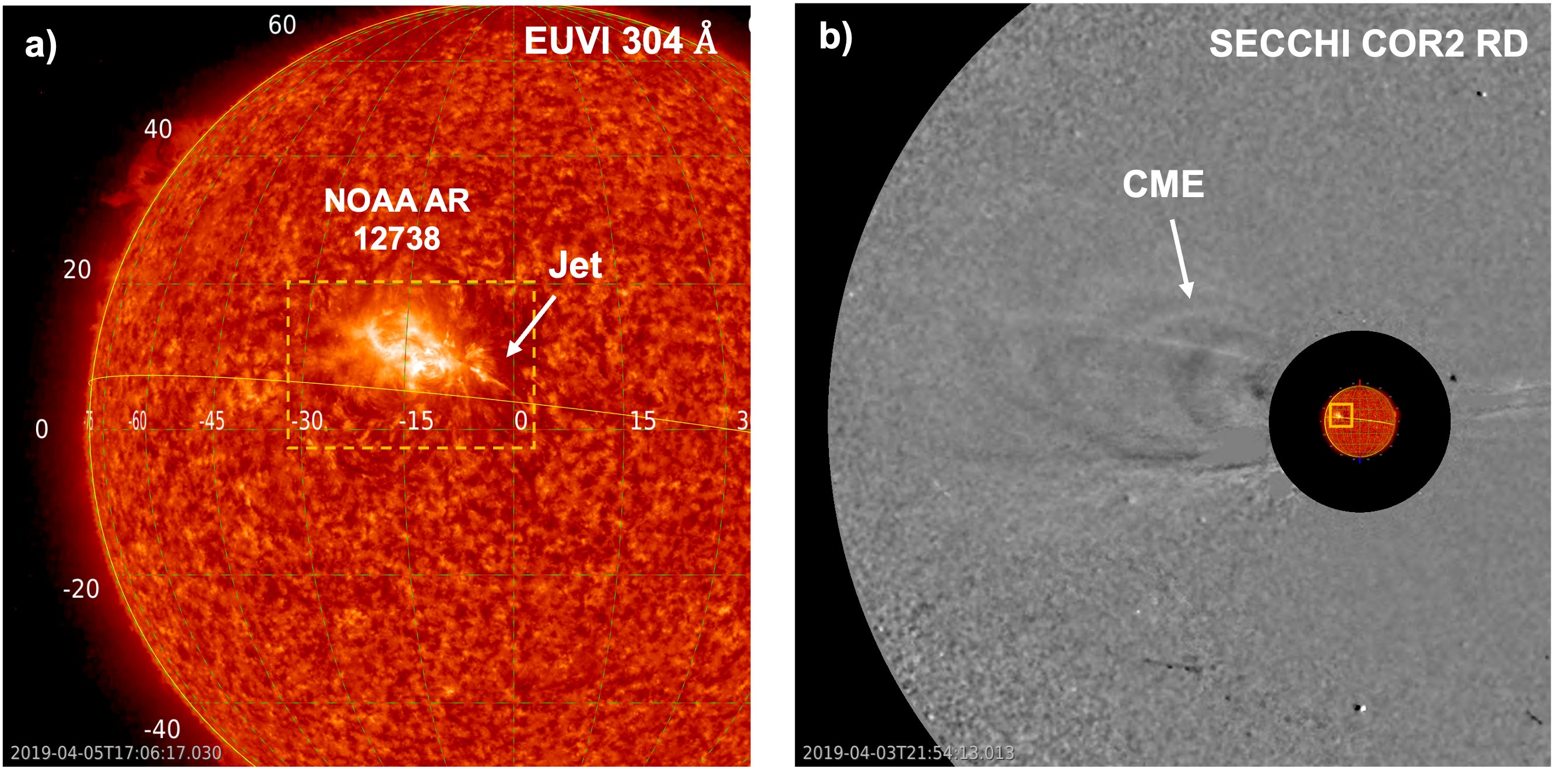}
     \caption{EUV image shows associated jet structure, while white-light observations capture one of the CMEs observed during the second PSP perihelion. Panel (a) displays a jet associated with Event 8 originating from the NOAA AR 12738, observed in the 304~\AA~wavelength channel of the EUVI instrument onboard STEREO-A. Panel (b) presents a combined image from 03 April 2019, featuring the EUVI 304~\AA~observation alongside white-light observations from the SECCHI COR2 instrument, highlighting a CME. Image created using Jhelioviewer \citep{Muller17}}
              \label{Fig: Jet}
    \end{figure*}

       \begin{figure*}
    \centering
    \includegraphics[width=18cm]{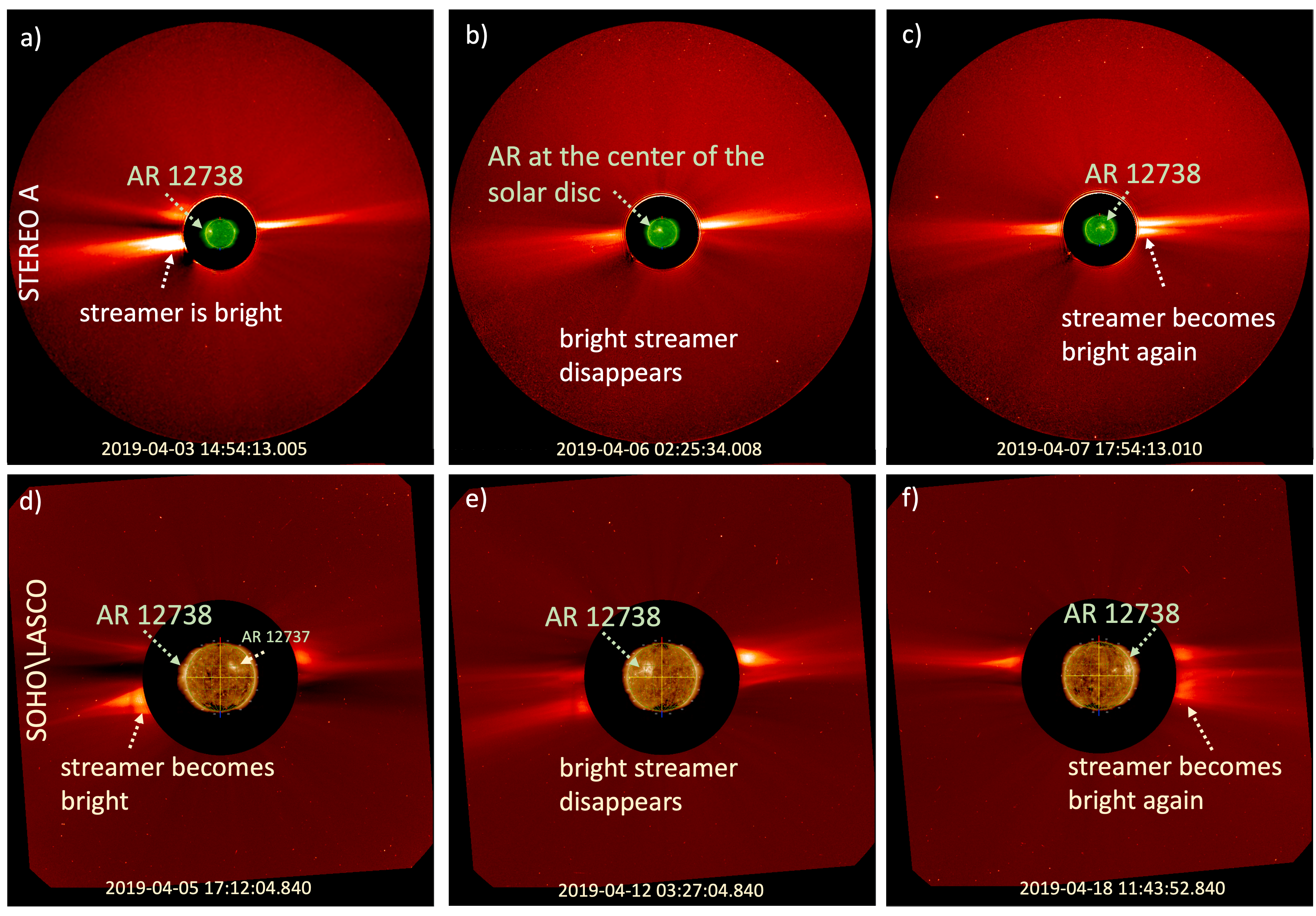}
     \caption{The white-light coronagraph observations, combined with EUV observations, show the streamers present during Event 8 and 9. The set of images highlights the streamer associated with AR 12738.  Panel (a to c) shows combination of SECHHI COR2 image and EUVI 195~$\AA$ images. Panel (d to f) shows combined AIA 193~$\AA$ and SOHO/LASCO C2 observations. Image created using Jhelioviewer \citep{Muller17}}
              \label{Fig: Coronagraph}
    \end{figure*}

\subsection{Event selection criteria}
\label{sec:selection}

We selected 11 type III radio bursts (\autoref{table1}) based on the following criteria:

\begin{enumerate}

    \item Radio bursts need to be observed by both Wind and STEREO-A to obtain dynamic spectra (see Fig.~\ref{Fig: Radio Spectra}) and direction-finding observations.

    \item Type III radio burst should be well distinguishable from other types of radio emission. We aim to have type III bursts as isolated as possible, in order to have a clear and not visibly overlapping frequency time profiles (see Fig.~\ref{Fig: Time Profiles}). We excluded type III bursts appearing in the big groups of numerous overlapping bursts. 
    
    \item Selected burst need to be of significantly higher flux intensity, approximately two orders of magnitude above the background level, in order to secure good triangulation results. 
\end{enumerate}

We note that most of the type III bursts occur within few minutes sometimes resulting in not completely simple time profiles (Fig.~\ref{Fig: Time Profiles}). This complexity in the time profiles can lead to potentially erroneous estimation of the radio source positions (Deshpande \& Magdaleni{\'c} in prep.).

The EUV observations showed that all selected type III radio bursts were associated with the active region (AR) NOAA 12738 (Fig.~\ref{Fig: Jet} and \ref{Fig: Coronagraph}). The source region was located at about 17$^\circ$ behind the east solar limb as seen from Earth on the day of the last observed type III bursts (05~April~2019). The AR was rather simple with $\alpha/\beta$ photospheric magnetic field configuration defined once it rotated on the visible side of the Sun. For all of the studied type III bursts, EUV observations were only available from STEREO-A. Observations at all of the 4 wavelength channels (see Sect~\ref{sec:Method}) provided by STEREO-A showed that the selected type III bursts were associated with small jet and jet-like structures (see example in Fig.~\ref{Fig: Jet}). 

The coronagraph observations showed few very faint and narrow CMEs during the considered time interval. The CMEs were mostly propagating along the streamer regions seen above the East solar limb (see Fig.~\ref{Fig: Coronagraph}). Propagation of the CMEs might have changed the magnetic field  configuration of the ambient solar corona and influenced the propagation of the type III electron beams (Magdaleni{\'c} et al. in prep.). We also note that the white light observations show the existence of a streamer that seem to originate close the NOAA AR 12738 (Fig.~\ref{Fig: Coronagraph}). The visual inspection of the EUV observations show streamer which initially appears bright when the AR is near the east limb (see Fig.~\ref{Fig: Coronagraph}, panels a and d). As the AR approaches the central meridian, the streamer becomes faint and almost not visible due to projection effect (Fig.~\ref{Fig: Coronagraph}, panels b and e). The streamer becomes bright again as the AR moves toward the west solar limb (Fig.~\ref{Fig: Coronagraph}, panels c and f).

\begin{figure*}
    \centering
    \includegraphics[width=18cm]{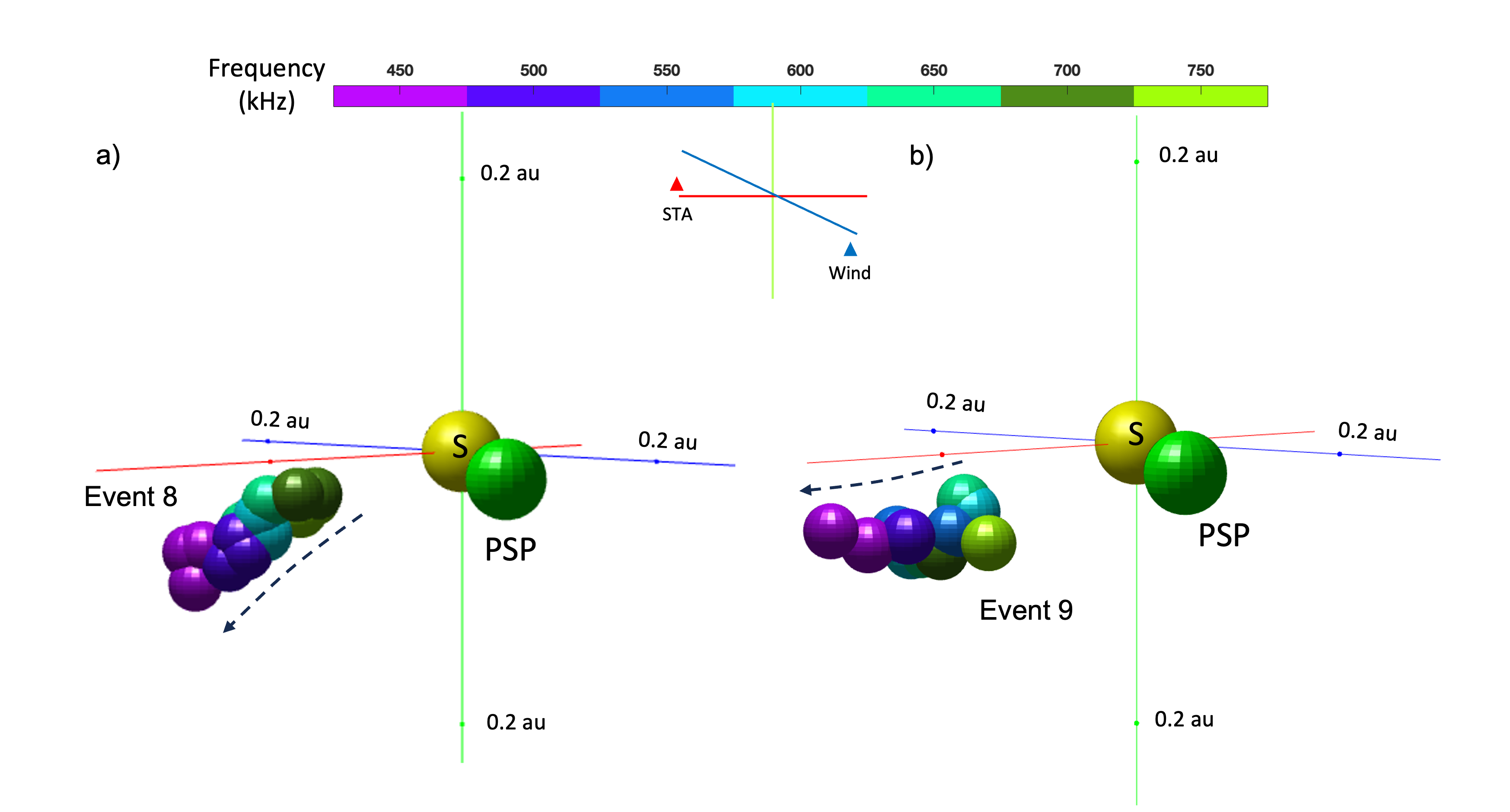}
     \caption{Radio source positions obtained using DF data in the 3D space (colorful spheres in blue to violet shades). The yellow and green sphere represent the Sun and PSP, respectively. The sizes of radio sources, Sun and PSP are not relative to each other.
     Panel (a) displays the radio source positions for the first type III bursts shown in the dynamic spectra (Fig.~\ref{Fig: Radio Spectra}). The radio source positions show the propagation path of the type III radio burst in the southward direction from the ecliptic plane. Panel (b) presents estimated radio source positions (obtained with radio triangulation) for the second type III burst observed at $\sim$17:15~UT in STEREO-A and Wind. The radio source positions show the propagation path of the type III radio burst more in the equatorial plane than in the case of the first type III.}
              \label{Fig: Radio Position}
    \end{figure*}

 \begin{figure*}
    \centering
    \includegraphics[width=18cm]{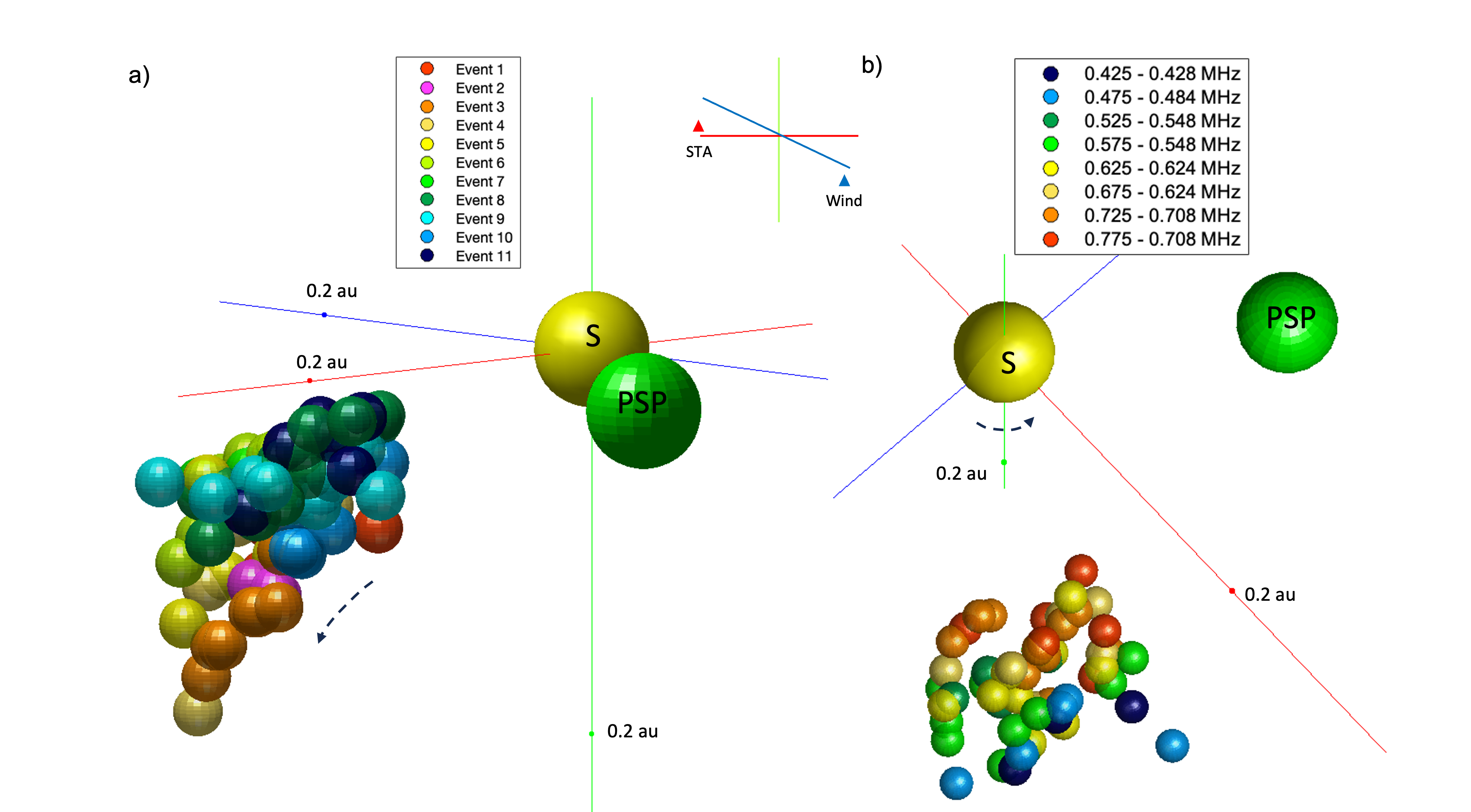}
     \caption{All type III radio bursts plotted together to visualize their propagation path relative to each other. Panel a) the propagation paths of all type III bursts presented in the similar viewing angle as the Events 8 and 9 (Fig.~\ref{Fig: Radio Position}). The spread of the sources indicates the rotation of the associated AR. Each bursts is presented in a different color. Panel b) Radio sources of all type III bursts plotted together but now color coded by the selected frequency pairs.}
     \label{Fig: all_typeIII}
    \end{figure*}

\subsection{Different methods of estimating coronal electron density}
\label{sec:densitymethod}

    \subsubsection{Radio density}
    
    The plasma frequency is due to collective oscillations of free electrons in the plasma. It is determined by the electron density (n), the elementary charge ($e$), and the electron mass ($m_{e}$). Mathematically, it can be expressed as, 
\begin{equation} \label{eq:2}
    f_p = \sqrt {\frac{ n~e^2}{\pi~\epsilon_{0}~m_e}}
\end{equation}

\noindent Electron density generally decreases with the increasing distance from the Sun, this makes the simultaneous decrease of the local plasma frequency. Such a behavior allows us to estimate the densities corresponding to each plasma frequency associated with the radio emission at the specific distances from the Sun. The radio triangulation method can be used to obtain the heights in the corona to which the estimated density corresponds. We will call density obtained in the described way ``radio density''.

\subsubsection{EUHFORIA density}

EUHFORIA simulates density, velocity, and magnetic field in the 3D space, all the way from the Sun to Earth and up to 2~au. In the framework of a single-fluid ideal-MHD model like EUHFORIA, the plasma is treated as quasi-neutral "fluid" with equal contribution of the electrons and protons which are not differentiated. Under this assumption, dividing the fluid number density by two gives a rough estimate of the electron (or ion) number density.

As the main input to EUHFORIA we used the synoptic GONG and ADAPT magnetograms (see Sect~\ref{sec:Method}). The source surface radii (SSR) of the PFSS, which defines height from which all the magnetic field lines are considered as open is combined with the source surface heights (SSH) of SCS model. In the default set up of EUHFORIA the SSH and SSR are considered to be 2.3 and 2.6~$R_\odot$ respectively. For modeling with GONG magnetograms we also used different pairs of the SSR values, in order to increase the accuracy of the background solar wind modeling. The solar wind speed and number density is described in the same way as in \citealp{Pomoell18} (eq. 2 \& 4).
In this study we used the intermediate resolution of EUHFORIA defined with angular resolution of $2^\circ$ with 512 grid cells in radial direction from 0.1~au to 2~au. We note that EUHFORIA allows also low ($4^\circ$ angular resolution with 256 grid cells) and high resolutions ($1^\circ$ angular resolution with 1024 grid cells).

With EUHFORIA we obtain simulation results in the 3D space which gives us the possibility to estimate the EUHFORIA electron density at any specific position. For that purpose we insert so-called virtual spacecrafts in the modeling domain \citep[see Fig.1 in][]{Scolini23, Valentino24}. Using virtual spacecraft allowed us to obtain EUHFORIA time series both at the radio source positions and along the PSP propagation trajectory.

\section{Case study of Event 8 and 9}
\label{sec:case}

 The case study of two type III radio bursts recorded on 05~April~2019 aims to demonstrate different methods for obtaining the solar wind densities and the ways to compare estimated values. The two selected type III bursts have peak intensities at 17:06~UT and 17:17~UT as detected by Wind, and around 17:05~UT and 17:16~UT as observed by STEREO-A. The peak times as observed by PSP are listed in Table~\ref{table1}. Similarly to all other type IIIs these two Events were sourced in the NOAA~AR~12738 and associated with jets. Due to the insufficient time resolution of the EUV data, it was challenging to determine which specific jet was responsible for each individual type III radio burst.

Fig.~\ref{Fig: Radio Spectra} shows the dynamic spectra, of Events 8 and 9. The dynamic spectra were recorded by Wind, STEREO-A and PSP covering the frequency range of 0.1 to 10~MHz. The Autoplot tool \citep{Faden10} was used for the dynamic spectra data plotting. Part of Event 8 is missing in the Wind spectrogram (between 17:14~UT to 17:19~UT) due to data gaps. All three spacecrafts observe the type III radio bursts as intense emissions isolated from each other (Fig.~\ref{Fig: Radio Spectra}). The maximum intensity of the radio emission at the DF frequency at 725 KHz, as observed by STEREO-A,  is $8.29 \times 10^{-17}$~W/m$^2$/Hz, which is about 100 times larger than at a similar frequency in Wind observations ($1 \times 10^{-19}$~W/m$^2$/Hz at 708 KHz). The intensity of the radio emission at the closest PSP frequency of 721 KHz for the same burst was $3.92 \times 10^{-14}$~W/m$^2$/Hz (Fig.~\ref{Fig: Radio Spectra}). A large difference between the STEREO-A and Wind type III intensity indicates that the burst likely propagates more towards STEREO-A (Fig.~\ref{Fig: position}) than towards Wind spacecraft. However, as PSP is at 0.16~au and STEREO-A and Wind are at about 1~au, direct comparison of the radio burst intensity and its directivity using all three viewing points is somewhat ambiguous. 

The frequency-time profiles of studied bursts, as observed by both STEREO-A and Wind data, are rather symmetric with a single maximum. This symmetry of both bursts is better seen in STEREO-A observations. We note that due to lower temporal resolution of the Wind data, the profile of Event 9 is likely not fully resolved (Fig.~\ref{Fig: Time Profiles}) while the time profile of Event 8 is affected by the data gap (Fig.~\ref{Fig: Radio Spectra}).

We note that it is difficult to state with full certainty whether the analyzed type III time profiles represent only single isolated bursts. Recent results by \citet[][see Fig.~6]{Gerekos24}, based on high temporal (1.43~ms) and frequency (7.41~KHz) resolution observations, suggest that more than one fully resolved type III profiles could be expected in the WIND and STEREO-A observations. The rather symmetric time profiles indicate that potentially unresolved type III bursts are temporally very close to each other.

Additionally, the dynamic spectra (Fig.~\ref{Fig: Radio Spectra}) reveal faint blob-like structures at the low-frequency part of both bursts (between 0.2 and 0.3~MHz), clearly observed by STEREO-A and PSP. No fine structures of the similar shape can be observed in the high frequency part of the bursts indicating that the blob-like structure are possibly different than the stria substructures previously reported \citep[e.g.,][]{Jebaraj22}. It is possible that the blob-like structures are observed in the STEREO-A and PSP data because the radio burst propagates towards STEREO-A, and close to the PSP. The reasons for not seeing the blob-like structure in the Wind data are possibly the radio burst propagation path away from Wind spacecraft and/or the instrumental effect. These structures observed in the dynamic spectra suggest that the measured time profiles at direction-finding frequencies are not simple but rather a convolution of at least two distinct components. All these characteristics of the two presented events indicate that the type III bursts likely consist of either: (a) both fundamental and harmonic components \citep{Jebaraj23a}, or (b) at least two closely spaced type III bursts \citep[e.g.,][]{Jebaraj22, Gerekos24}. We note that both of mentioned studies which were able to resolve the bursts time profiles considered type III bursts at frequencies higher than in this study namely above 1~MHz \citep[see Fig.~1 in][]{Jebaraj23b} and 13.5~MHz \citep{Gerekos24}.
Even if the time profiles of the bursts are observed with high temporal resolution as in the study by \cite{Vecchio24} who considered type IIIs observed at frequencies above $\sim$ 3~MHz, at frequencies below 1 MHz it is likely that time profiles will be convolutions of fundamental and harmonic emission or subsequent bursts \citep[see Fig.~9 in][]{Dulk84}. Therefore, increasing temporal resolution further will often not allow clear deconvolution of the overlapping components, in particular below the herein considered frequencies below 800 kHz.

  \begin{figure*}
    \centering
    \includegraphics[width=18cm]{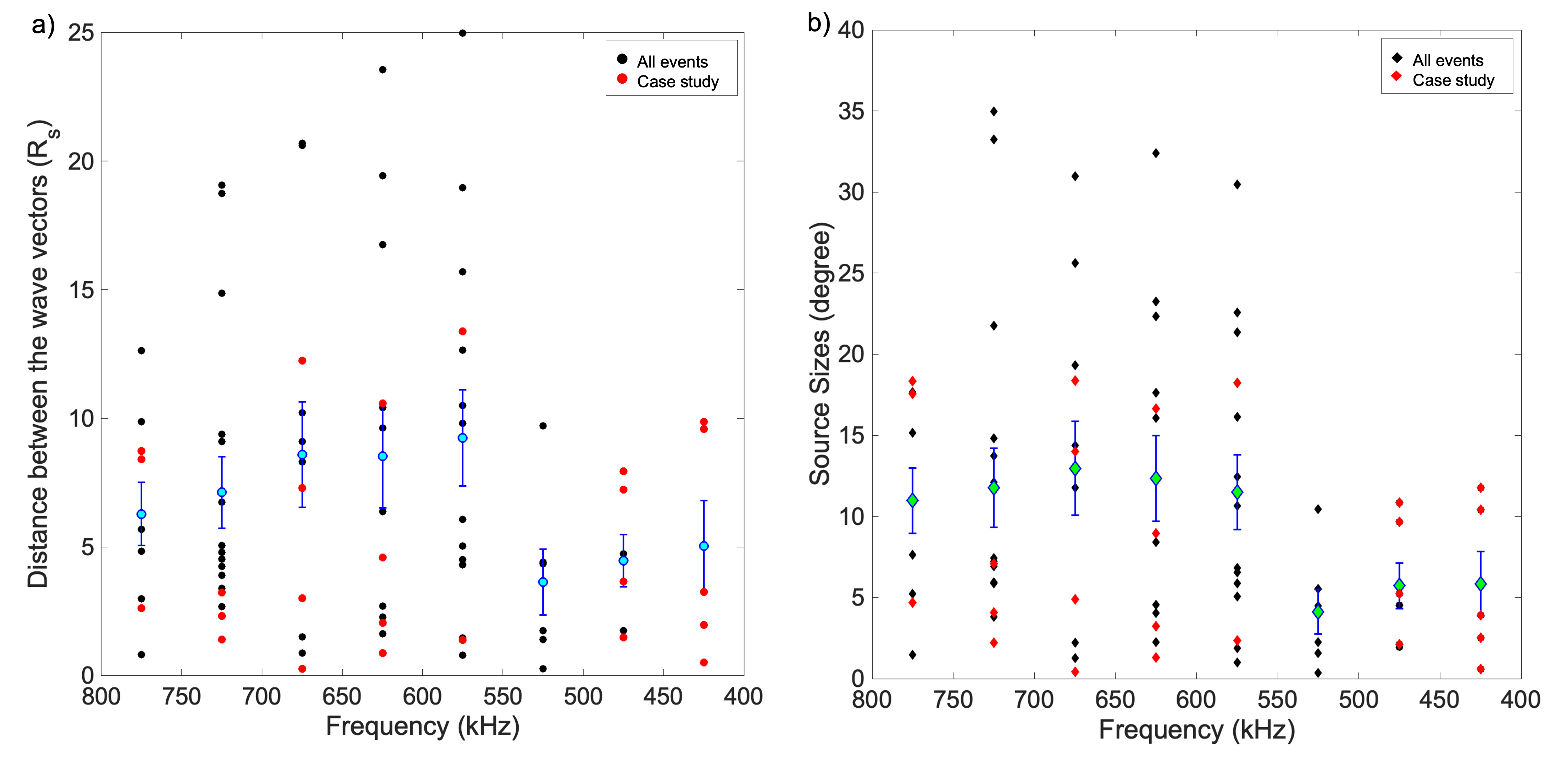}
     \caption{The variation of radio source sizes with frequency for all type III radio bursts studied herein. The red points represent the case study and black points shows all the selected events. The blue error bars plotted on the data represent the standard deviation estimated from bootstrapping while the cyan colored points represent the mean calculated using the bootstrapping method.  Panel (a) shows the radio sources in terms of the distance between the wave vectors. The average radio source size is up to $5~R_{\odot}$, while panel (b) shows source sizes in terms of degrees.}
     \label{Fig: distances}
    \end{figure*}

\section{Triangulation results}
\label{sec:triangresults}

In this section we discuss triangulation results for the Events 8 and 9 (see Fig.~\ref{Fig: Radio Position}), and position these results in context with the statistics of the whole data set.

Using radio triangulation with Wind \& STEREO-A data we obtained the source positions of the type III bursts for the seven selected frequency pairs (see Sect.~\ref{sec:triangulation}). The obtained radio source positions, shown in Fig.~\ref{Fig: Radio Position}, are at heights range 15 to $60~R_{\odot}$. The spheres marking the radio sources in Fig.~\ref{Fig: Radio Position} have a fixed size of 0.02~au for all frequency pairs, regardless of the actual distance between wave vectors determined by the direction-finding method. This approach is chosen to improve the clarity of the radio sources' propagation direction, similar to the method used by \cite{Magdalenic14}. Varying sphere sizes at different frequency pairs (as shown in Fig.~\ref{Fig: distances}) can visually distort the apparent propagation path of the bursts. Figure~\ref{Fig: Radio Position}a demonstrates that the radio sources of Event 8 are well-aligned, clearly showing a propagation path directed southward from the ecliptic plane. In contrast, the radio source positions for Event 9 are less coherent, revealing a propagation path nearly parallel to the ecliptic plane (Fig.~\ref{Fig: Radio Position}b). This indicates that two consecutive type III bursts, separated by only about 10 minutes, can have significantly different propagation directions and originate from fast electron beams traveling along distinct, although not necessarily widely separated, magnetic field lines.

We also plotted all type III radio bursts together to visualize their propagation paths relative to each other (Fig.~\ref{Fig: all_typeIII}a). Most type III bursts propagate southward from the solar ecliptic plane, possibly due to changes in magnetic field configuration caused by the CMEs and streamers observed in white-light data (Section~\ref{sec:Method}). We note that even small changes in the reconnection site (source of the type III electron beams) or the underlying magnetic field configuration can result in electrons accessing different magnetic flux tubes, leading to slightly different propagation paths and possibly somewhat different apparent radio source locations. Figure~\ref{Fig: all_typeIII}b shows all type III bursts, with radio sources color-coded by frequency. Higher-frequency sources (red spheres) are generally located closer to the Sun, while lower-frequency sources (blue spheres) appear farther away. This figure also highlights the spread of type III events as the AR rotates. For example shown in Fig.~\ref{Fig: Radio Position} two type III bursts are separated by less than 15 minutes and during this interval the Sun rotates about 0.13~deg. Taking this into account, even if the propagation path of the type III bursts would be exactly along the same magnetic field line the impact on the estimated radio source propagation would be small. The time difference between the first and the last type III bursts is about 4~days which accounts for about 52~deg in the change of the AR position. The spread of the apparent source positions shown in the Fig.~\ref{Fig: all_typeIII}b is about 40-50~degrees which is in accordance with the AR rotation.

The distances between wave vectors ($d$), obtained through radio triangulation, and the estimated radio source sizes ($\theta$) (Section~\ref{sec:selection}) are shown in Fig.~\ref{Fig: distances} as functions of the STEREO-A observing frequency (the first frequency in each triangulation pair).According to eq.~\ref{eq:1}, radio source size depends on the distance between wave vectors and the source-to-Sun distance, resulting in variations in data point distributions between the two panels in Fig.~\ref{Fig: distances}. Red symbols represent results for Events 8 and 9, while black symbols indicate all other studied events. The radio source position falls within the distances between wave vectors. We therefore consider that distance between the wave vectors represents the upper limits for source sizes determined by the triangulation method.

Figure~\ref{Fig: distances} shows significant spreads in values of both $d$ and $\theta$ at each frequency. Although most data points cluster at lower values, some large values for $d$ and $\theta$ were observed, making a simple mean inappropriate. To address this, we applied a bootstrap method \citep{Dissauer2018} to estimate the mean source sizes and their uncertainties more accurately. Bootstrapping involves repeatedly resampling the data (with replacement) to estimate statistical distributions without assuming underlying data distributions. We performed 10,000 bootstrap resamples, computing means and standard deviations for each set. This method provides robust estimates of uncertainties, reflecting the data's inherent variability. Error bars derived from bootstrapping are shown in Fig.~\ref{Fig: distances}.

The bootstrap-derived mean values of $d$ and $\theta$ show only a slight increase in the distance between wave vectors at frequencies from 800 to 550~kHz, with small effect on source sizes. Mean values for the next lower frequency range (550 to 400~kHz) are unexpectedly smaller. We note that not all type III bursts go to such low frequencies and often even if they do, the intensity of the radio emission is very low which does not allow us to perform accurate radio triangulation. Therefore, we have less data points below 550 kHz which could influence the obtained source sizes. Considering the full frequency range (800--400~kHz), we find no clear systematic variations in source sizes i.e., systematic increase in $d$ or $\theta$ at lower frequencies was not found contrary to recent results in metric-range \citep{Saint-Hilaire13, Gordovskyy22}. This indicates that the scattering effects are probably stronger in the low corona than at larger distances from the Sun. It is important to note that this lack of the systematic source size variation with the frequency is not influenced with the source size definition. Namely, we employ the same source size definition as in previous case studies \citep{Magdalenic14, Jebaraj20} which found somewhat more systematic variation with the frequency decrease. The larger number of considered type III bursts in this study provides more general conclusion.

\begin{figure*}
    \centering
    \includegraphics[width=18cm]{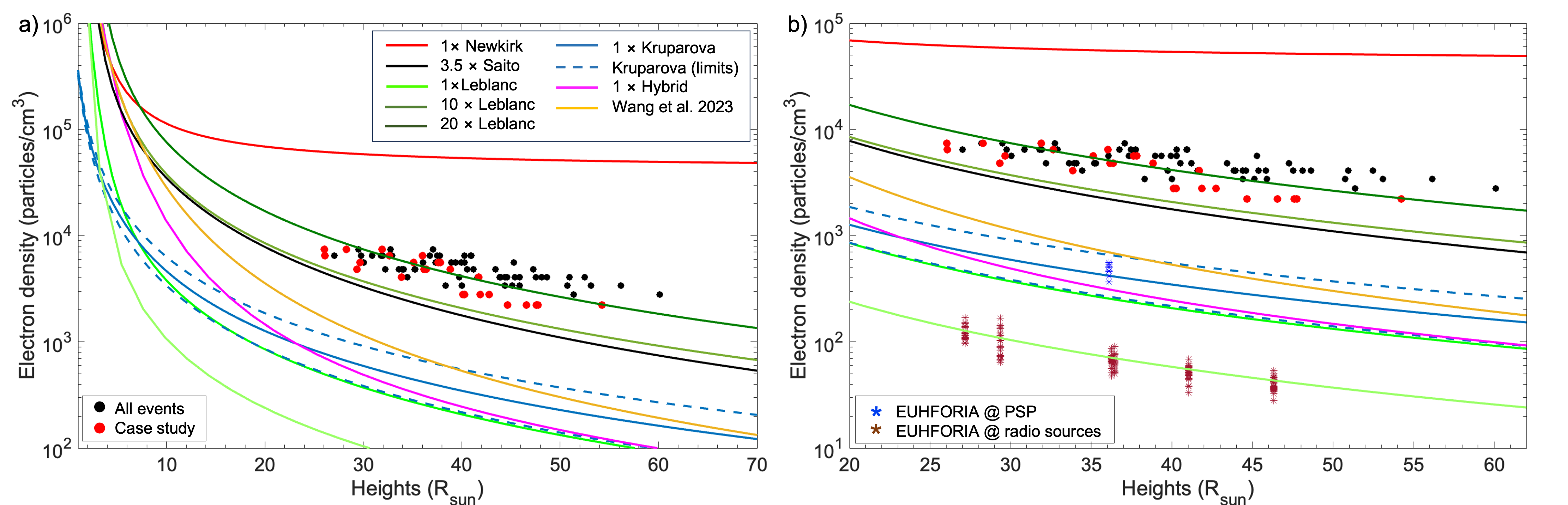}
     \caption{Comparison of the densities obtained in this study with some previous observational results and models. The electron density models such as, Newkirk \citep{Newkirk61}, Saito \citep{Saito70}, Leblanc \citep{Leblanc95}, Hybrid \citep{Vrsnak04} and \cite{Wang23}, Kruparova model \citep{Kruparova23} are shown with red,  black, shades of green, magenta, yellow and blue lines respectively. Red points represent the radio densities obtained from the case studies, while black points represents all events. Panel b) shows EUHFORIA densities obtained at PSP and radio source positions with all realizations of ADPAT magnetic maps in comparison with other density models and radio densities}
              \label{Fig: Density model}
    \end{figure*}

\section{Comparison of the electron densities obtained with different methods}
\label{sec:comparison}

\subsection{Radio densities}
\label{sec:radiodensity}

After obtaining the 3D positions at which the radio emission was generated in the corona, using eq.~\ref{eq:2} and knowing the frequencies at which radio burst was studied, we estimated the radio densities. Employing different frequency pairs (see Section~\ref{sec:triangulation}) allowed us to map the radio densities along the propagation path of the type III electron beams in the 3D space (Fig.~\ref{Fig: Radio Position}). Additionally, in order to directly compare radio densities with the generally used 1D coronal electron density models we converted the 3D information of the radio source position to 1D heights (see Section~\ref{sec:densitymethod}). Results of this part of the study are presented in Fig.~\ref{Fig: Density model}a which shows the density profiles as a function of height above the solar surface. The radio density obtained for Events 8 and 9 is marked with red circles, and for all other studied bursts with black circles. The generally used 1D coronal electron density models and presented with different colored lines. We find the radio densities obtained in this study to be situated systematically above the 10$\times$Leblanc or 3.5$\times$Saito model and along 20$\times$Leblanc model \citep{Leblanc95}. The obtained radio density values are somewhat larger than the usually estimated ones. However, the trend of the density change with the height closely follows the one of Leblanc model, which is to be expected as the Leblanc model was constructed based on the type III radio bursts.
Further, radio densities are also significantly higher than the density observed by PSP during its first ten close encounters, and reported in the statistical study by \cite{Kruparova23}, shown in the Fig.~\ref{Fig: Density model} with the blue lines. We note that the red points which represent results for the case study follow the same trend as the rest of the obtained radio densities, and they are situated just above the 10$\times$Leblanc density profile.

We note that studies employing very large densities to explain radio observations were also reported earlier, in both metric and kilometric range and amounted to as much as 10$\times$ Saito, 8$\times$ Newkirk etc. \citep[see e.g][]{Pohjolainen08, Jebaraj20}.

\subsection{EUHFORIA and PSP densities}
\label{sec:pspdensity}

In this section we discuss how ``EUHFORIA densities'' were obtained from EUHFORIA simulations \citep{Pomoell18} and its comparison with the radio densities (Section~\ref{sec:radiodensity}).

We employed the default set up of EUHFORIA code (Section~\ref{sec:Method}) to obtain EUHFORIA densities that can be compared with the radio densities. As the main input to EUHFORIA, we used GONG and ADAPT synoptic magnetograms at 03:00:00~UT on 05~April~2019. Since there were no CMEs directly temporally associated with the studied type III bursts, we use only the background solar wind model, without inserting CMEs.

Combining observations and the results of the radio triangulation study shows that the type III radio bursts propagate in the vicinity of PSP but do not exactly cross PSP path. That makes the direct comparison of the radio density and the PSP density difficult. Therefore, we make a two-step comparison, which includes EUHFORIA density and PSP density in the first step, and EUHFORIA density and radio density in the second step. However, to understand how accurate the solar wind density modeled with EUHFORIA is, we start with a comparison of the EUHFORIA modeling results with the PSP densities. 

Therefore, in order to compare radio density, as direct as possible, we used the 3D radio source positions obtained with radio triangulation as the artificial spacecrafts in the EUHFORIA's heliospheric domain \citep[see sketch in Fig.~\ref{Fig: position}b and also see Fig.1 in][]{Valentino24}. That allowed us direct comparison of the EUHFORIA density and radio densities along the propagation path of type III electron beams. 

Similarly to the method for comparison of modeling results and radio densities we placed virtual spacecraft at PSP positions. That allowed us to examine the time series observed by PSP and the one modeled by EUHFORIA at PSP position (Fig.~\ref{Fig: Radio Position}).

\begin{figure*}
    \centering
    \includegraphics[width=19cm]{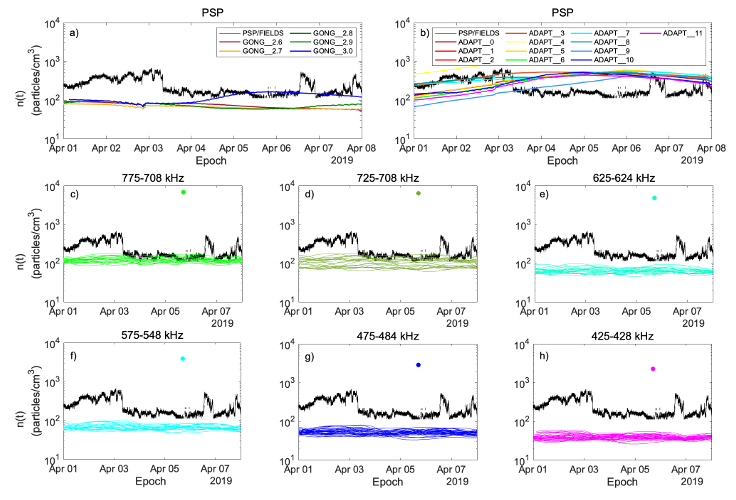}
     \caption{The comparison of PSP/FIELDS in situ electron density (black line) with EUHFORIA simulation results at various positions in the heliosphere. Panel (a) displays simulations at PSP position using GONG magnetograms, with different colored lines representing simulations performed with varying SSRs. The simulation using SSR = 3 $R_{\odot}$ shows the best agreement with the in situ data. Panel (b) shows simulations using default PFSS height at PSP position using all realizations of ADAPT magnetic maps, which in this case provide a better match than those using GONG magnetograms. Panels (c) to (h) present EUHFORIA time series obtained using ADPAT maps at radio source positions for different triangulation frequency pairs, with colored points indicating the electron density at the respective radio frequencies (radio density). Radio densities are notably higher than the in situ density, while the EUHFORIA results obtained at the radio source positions are underestimated.}
              \label{Fig: EUHFORIA_Time_series}
    \end{figure*}

\begin{figure}[ht!]
    \centering
    \includegraphics[width=9cm]{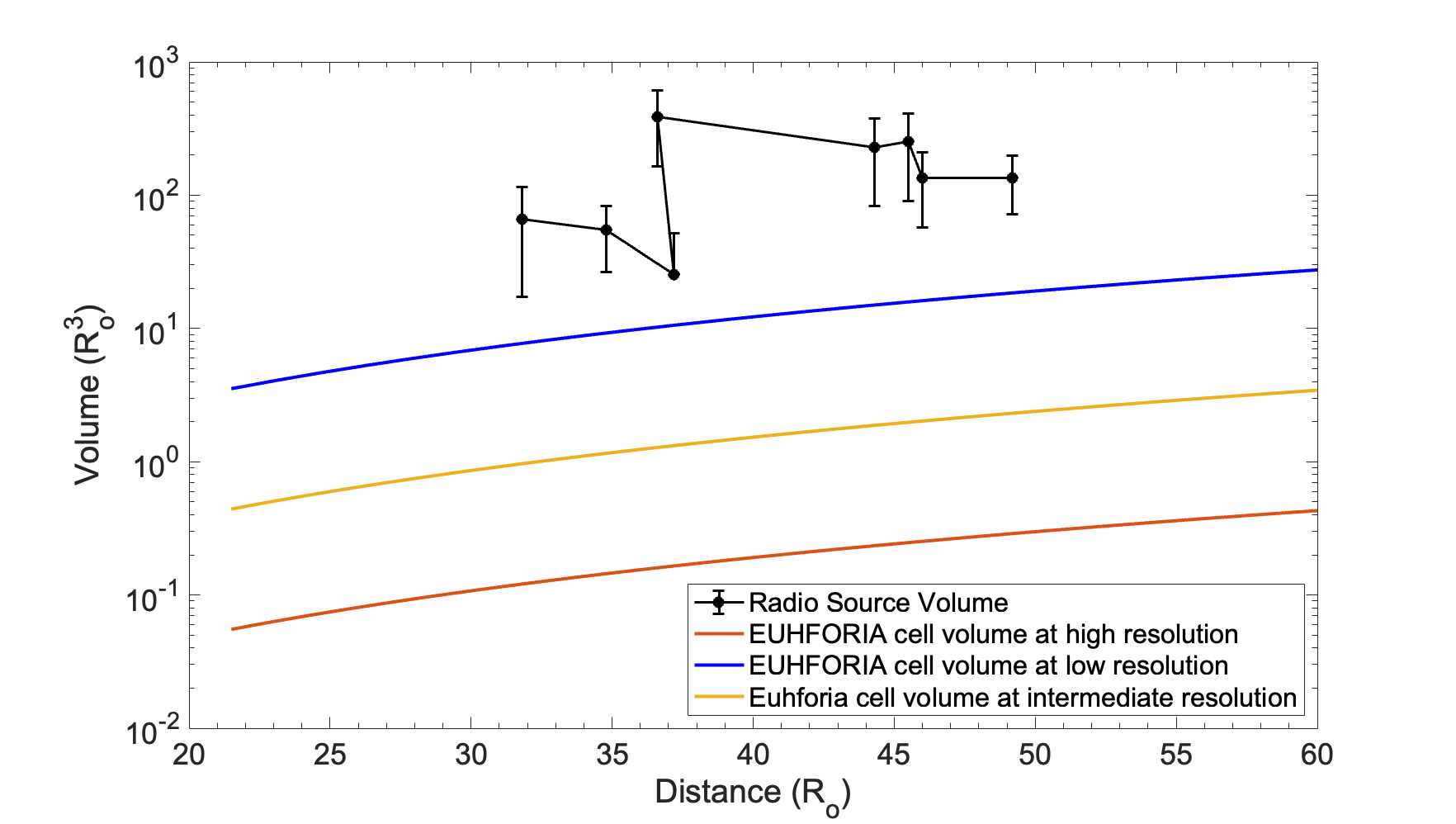}
     \caption{The variation of EUHFORIA cells volumes and radio source volume (black line) as a function of their distance from the Sun. The radio source volume points are estimated assuming spherical shape sources, with the distance between the wave vectors being sphere diameter. The blue, yellow, and red lines represent EUHFORIA cell volumes at low, intermediate, and high resolutions, respectively. The EUHFORIA cell volumes are notably smaller compared to the radio source volumes and therefore, can effectively model the regions occupied by the radio sources.}
              \label{Cell_size}
    \end{figure}

We initially utilized GONG magnetic maps to estimate inner boundary conditions for EUHFORIA (Fig.~\ref{Fig: EUHFORIA_Time_series}, panel (a)). However, we observed that the estimated density at PSP position was significantly underestimated when using the default PFSS height of 2.6~$R_\odot$. In order to test if the change of the SSR can improve the modeling accuracy \citep[as shown by,][]{Reville15, Asvestari19, Boe20}, additionally to the default set up where SSR=2.6~$R_\odot$ we also perform simulations for 4 different SSRs (2.7, 2.8, 2.9, 3.0~$R_\odot$). Our results, presented in Fig.~\ref{Fig: EUHFORIA_Time_series}, show the significant discrepancy between modeled and observed electron density at the beginning of the considered time interval when the PSP density is rather high, i.e., 00~UT on 02~April until the 12~UT on the 03~April. Starting from around 10~UT on 03~April the PSP density sharply decreases and then remains at that low value until the end of the considered interval. This low level of the PSP density was better reproduced by EUHFORIA than the density at the beginning of the considered period. During the interval of low PSP density its variability is between 100 to 200~particles/cm$^3$ which is somewhat higher than the variability we obtain with the modeling results (about 80~particles/cm$^3$) using SSR between 2.6 to 2.9 R$_\odot$ (Fig.~\ref{Fig: EUHFORIA_Time_series}). The modeling results with SSR = 3.0~R$_\odot$ show significant density increase, going to the values of up to 200~particles/cm$^3$, matching the PSP density. We need to note that in general, being the MHD model EUHFORIA is not able to reproduce the fast density-fluctuations like e.g. the sudden density increase observed by PSP at the end of the day on 06~April.

Since the modeling results obtained with the GONG magnetograms systematically underestimate the in situ density values, in order to try to improve the modeling accuracy we employed also ADAPT magnetograms. We run simulations using all 12 realizations of ADAPT magnetograms and the default PFSS height of 2.6~$R_\odot$. These simulations produced higher densities at PSP position during the first time interval 00~UT on 02~April until the 12~UT on the 03~April, aligning more accurately with PSP in situ data, as seen in Fig.~\ref{Fig: EUHFORIA_Time_series}b. However, in the second part of the considered time window 12~UT on 03~April until 08~April EUHFORIA with the ADAPT magnetograms input generates the overestimated density. The results from all ADAPT realizations differ from each other, yet all of them show a similar maximum density values of approximately 500~particles/cm$^3$. We note that the level of the variability between results obtained using different ADAPT realizations is larger during the intervals of the high density mapped by PSP (first mentioned interval). The ADAPT maps provide more intervals of better reconstructed in situ density, in comparison to GONG maps which result in the systematic density underestimation. Therefore, we consider the modeling with the ADAPT maps to be more accurate one and will use the best ADAPT realization 03 for modeling radio densities. We would also like to note that modeling coronal plasma characteristics at close to the Sun distances is quite difficult as the modeling accuracy depends on the various different factors \citep[][]{Samara24}.

GONG and ADAPT magnetograms are differently constructed. GONG magnetograms are purely observational, limited to the visible disk of the Sun as seen from Earth. In contrast, ADAPT magnetic maps are generated using a data assimilation model that combines high-resolution observational data (taken by SDO) with flux transport models to provide a time-evolving, full-surface representation of the Sun’s magnetic field, including regions not directly observed. This complex approach provides possibility for more accurate and globally consistent solar magnetic environment when ADAPT maps are input to EUHFORIA. Similarly to \cite{Perri23}, we obtained somewhat improved prediction of solar wind dynamics and heliospheric conditions when ADAPT maps were used. However, in order to confirm that ADAPT maps give systemically better modeling results a statistical study is necessary.
 
Panels c) to h) of Fig.~\ref{Fig: EUHFORIA_Time_series} show comparison of: (a) the electron density obtained with the radio observations (colorful points); (b) the PSP density; and (c) the EUHFORIA density at the radio source positions obtained using ADAPT maps as an input (colorful lines). The simulation results for the radio source positions obtained with the first frequency pair 775-708~KHz is quite similar to PSP electron density. As we go to the lower frequencies (the increasing heights of the radio sources), the EUHFORIA density at the radio source positions is systematically decreasing. Such a behavior is expected as the solar wind density generally decreases with the increasing distance from the Sun. We note that for the last frequency pair the difference becomes almost an order of magnitude. The EUHFORIA results obtained using ADPAT maps at PSP position are along the \cite{Kruparova23} model, which represents PSP densities. Similar discrepancy between differently obtained density is also seen in the Fig.~\ref{Fig: Density model}b where the EUHFORIA results obtained at PSP  and radio source positions (blue and brown stars) are compared with the different 1D density models and radio densities.
There are number of factors that induce the discrepancy of the coronal electron density seen in \ref{Fig: Density model}b \& Fig.~\ref{Fig: EUHFORIA_Time_series} and they will be discussed in details in Section~\ref{sec:uncertainities}.

In order to check how the angular and radial resolution influence the accuracy of the obtained simulation results in comparison to the radio source sizes (see Section~\ref{sec:densitymethod}) we estimated the EUHFORIA cell volume at different distances from the Sun. The estimation of the radio sources volume was done assuming sources are of a spherical shape, with the distance between the wave vectors considered as a sphere diameter. We obtained the error bars and mean value for the radio source volume from the bootstrapping method (Fig.~\ref{Fig: distances}a). In Fig.~\ref{Cell_size} we plot the obtained radio source volume as a function of the distance from the Sun estimated from the radio triangulation. The EUHFORIA cell volume is calculated as follows,

\begin{equation}
    dV = r^2 . \sin(\theta)~.~\Delta{r} . \Delta\theta . \Delta\phi 
\end{equation}

\noindent where, r is radial distance between the inner boundary of EUHFORIA and the considered cell, $\Delta{r}$, $\Delta\theta$, $\Delta\phi$ are the radial, latitudinal and longitudinal resolution of the EUHFORIA. Fig.~\ref{Cell_size} shows the comparison between EUHFORIA volume, in blue, yellow and red depending on the EUHFORIA's resolution, and radio source volumes (black circles). Although, the EUHFORIA cell volumes are smaller than the radio source volumes for all the resolutions, the cell volumes for the lowest EUHFORIA resolution are comparable to the radio source volumes. Therefore, selection of the intermediate resolution of EUHFORIA is justified as it is capable of fully capturing the radio variations of the density for different closely spaced radio source positions. The jump of the volume at about 37 $R_\odot$ is due to the larger spread of the distances between the wave vectors from the triangulation results as around 650-550 kHz as can be seen in it Fig.~\ref{Fig: distances}a.

\subsection{Density fluctuations in PSP data}
\label{sec:fluctuation}
Density fluctuations in the solar corona can influence the observed radio source positions \citep{Kontar19, Krupar20, Jebaraj20, Sishtla23}.
In order to understand how strong is the influence of density fluctuations on the estimated radio source positions in the case of studied type III bursts, we analyzed fluctuations of the PSP electron density. 

With the aim to inspect density fluctuations during longer time window and obtain more general estimate we considered the full duration of the second PSP close approach covering the time interval from 01~April up to 08~April. This time interval covers all selected type III radio bursts (observed between 02~April  and 05~April). In order to estimate the level of the small scale fluctuations, we selected a time scale of one minute during which PSP encompasses distances which are short in comparison to the radio source sizes as presented in Fig.~\ref{Fig: distances}. We note that the temporal resolution of FIELDS/RFS time series is 7~s, while the full duration of the type III bursts at a considered DF frequencies is about 10~min (Fig~\ref{Fig: Radio Spectra}). Therefore, looking at 1~min time scale fluctuations seems to be reasonable assumption. We emphasize that the considered time interval of the PSP observations is 8 days, during which PSP maps the solar wind plasma characteristics at distances of about \mbox{34\,--\,40} $R_\odot$ which is the region in which the majority of the radio triangulation points are located (see Fig.~\ref{Fig: Density model}). The density fluctuations estimated in such a way have also the statistical weight, as we sample large portion of the ambient plasma characteristic. 
As such the PSP observations are the best proxy possible for estimation of the density fluctuations, even if the PSP is not exactly at the same position as the radio sources. Other ways of estimating density fluctuations e.g. using the remote sensing observations or theoretical estimations \citep{Thejappa92, Krupar18} are expected to be less accurate.

    \begin{figure}[ht!]
    \centering
    \includegraphics[width=9cm]{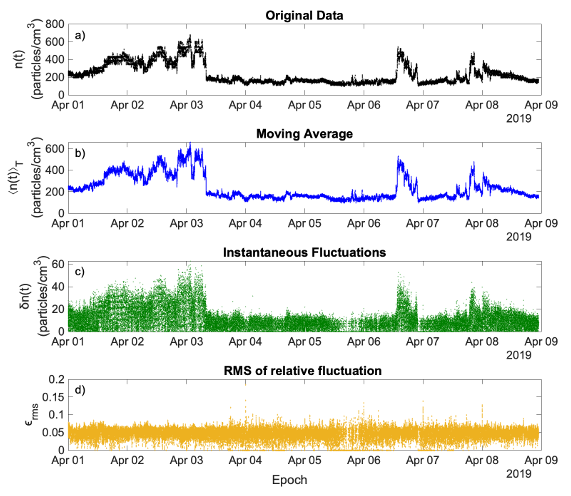}
     \caption{Panel (a) displays the electron density measured by PSP/FIELDS instrument. Panel (b) presents moving averages of the electron density over a time window of approximately one minute. In panel (c), we illustrate the maximum small-scale fluctuations for this time window within the specified epoch. Finally, panel (d) depicts the RMS relative density fluctuations observed. Notably, the instantaneous density fluctuations tend to be more pronounced in regions of higher electron density.}
              \label{Fig: small_scale_fluctuations}
    \end{figure}

    \begin{figure}[ht!]
    \centering
    \includegraphics[width=9cm]{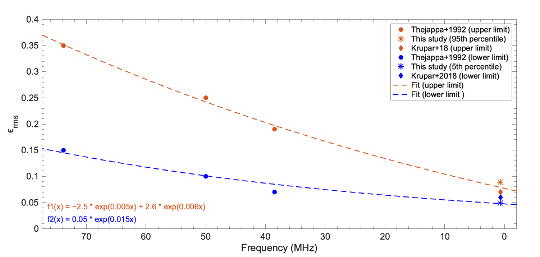}
     \caption{The RMS relative density fluctuations obtained using a 1-minute time window from PSP in situ data are shown as magenta stars. The two data points represent the 5th and 95th percentiles of the obtained values (Fig.~\ref{Fig: small_scale_fluctuations}d). These results are then compared with those obtained by \citet{Thejappa92} and \citet{Krupar18} (blue and orange circles and diamonds, representing the lower and upper limits, respectively). A fit is applied to both datasets to highlight the good agreement between them.}
              \label{Fig: epsilon}
    \end{figure}

\begin{figure*}
    \centering
    \includegraphics[width=18cm]{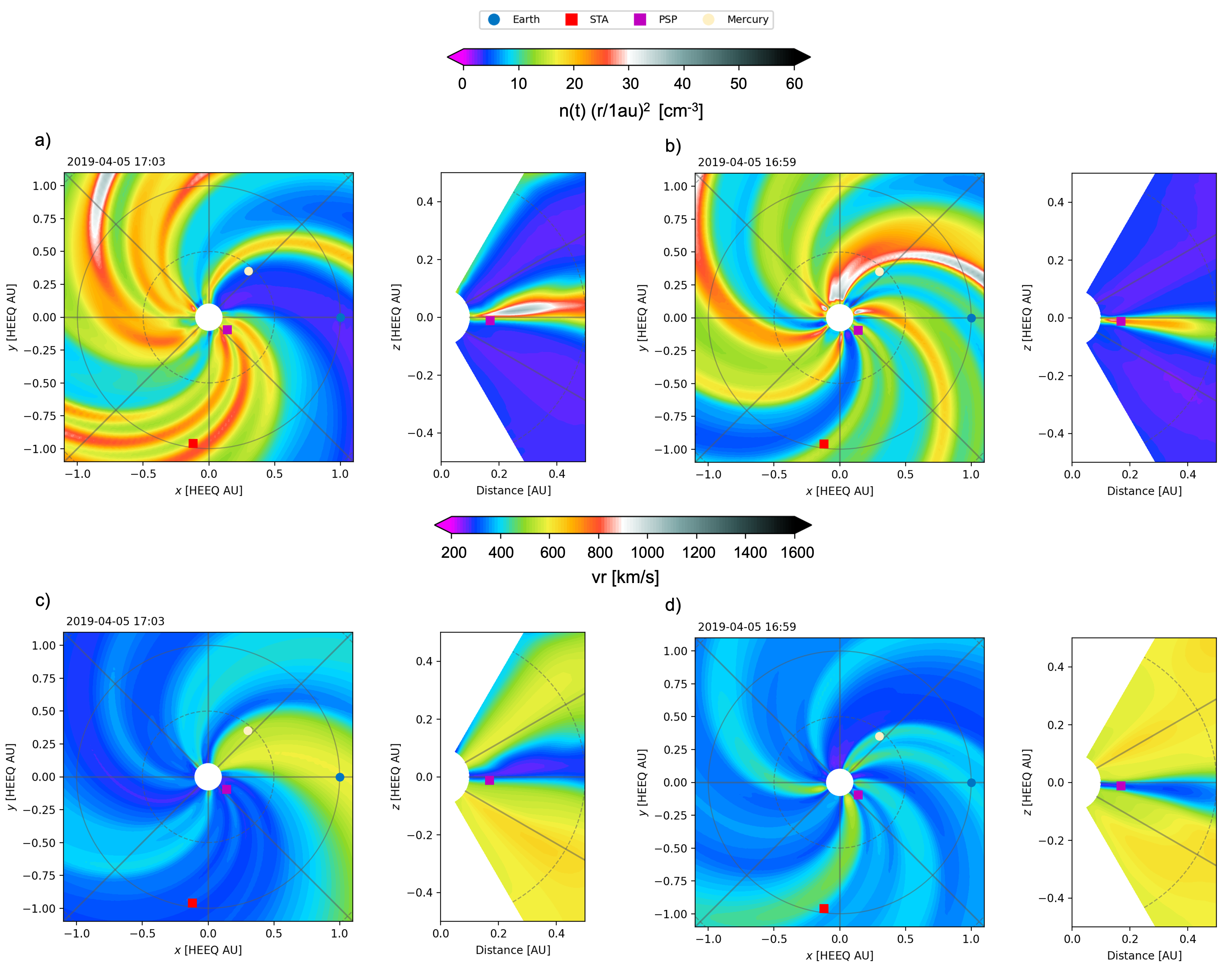}
     \caption{Equatorial and meridional slice of the EUHFORIA simulation domain. The meridional plane contains PSP. In each panel, the left plot depicts the equatorial plane of the heliospheric domain, while the right plot shows the meridional view of the same domain. Panels a) and c) display 2D density maps obtained using GONG and ADAPT magnetic maps, respectively. Panels b) and d) illustrate 2D velocity maps using GONG and ADAPT magnetic maps, respectively. The meridional plots highlight PSP's trajectory through a high-density, low-velocity region in both simulations with GONG and ADAPT magnetograms.}
              \label{Fig: EUHFORIA2D}
    \end{figure*}

We calculate the centered moving averages of the PSP density $\langle n(t)\rangle_T$, by sliding the time window of  1~minute along the full considered epoch (Fig.~\ref{Fig: small_scale_fluctuations}b). The maximum fluctuations $\delta$n(t) were then estimated as the difference between the original data points n(t) and the moving average for the 1~min time window (Fig.~\ref{Fig: small_scale_fluctuations}c). The instantaneous fluctuations $\epsilon$(t) were calculated as follows:

\begin{equation}
    \epsilon(t) = \frac{\delta n(t)}{\langle n(t)\rangle_T}.
\end{equation}

Root mean square (RMS) of the relative fluctuations ($\epsilon_{rms}$) can be calculated as,

\begin{equation}
    \epsilon_{rms} = \sqrt{\frac{\langle \delta n(t)^2\rangle_T}{\langle n(t)\rangle^2_T}}
\end{equation}

Our results show that the $\epsilon_{rms}$ throughout the given epoch were about 0.05 (Fig.~\ref{Fig: small_scale_fluctuations}d). If we consider longer time window of 10~min for estimating moving averages, which will then correspond to a larger scales and it will encompass the whole duration of the type III bursts, we obtain RMS density fluctuations on average at the level of about 0.06. We note that even considering such a longer time window, not really suitable for the study of short duration type III bursts, the obtained level of density fluctuations remains as small as 6 percent. 

We compared obtained values of $\epsilon_{rms}$ to the one found by \citet{Thejappa92} in the metric range (Fig.~\ref{Fig: epsilon}). These authors obtained, at close to the Sun distances for frequencies of about 38.5, 50 and 73.8~MHz, the density fluctuations of \mbox{0.07\,--\,0.19}, \mbox{0.1\,--\,0.25}, \mbox{0.15\,--\,0.35}, respectively. Employing Monte Carlo simulation technique \citet{Krupar18} obtained \mbox{0.06\,--\,0.07} of relative electron density fluctuations at the distances between 8 and $45~R_{\odot}$. We obtained the 5th and 95th percentiles of the $\epsilon_{\text{rms}}$ to determine the upper and lower limits of our results. We then compared these values with those obtained by \citet{Thejappa92}. The fit showed good agreement between the metric range results and our findings (Fig.~\ref{Fig: epsilon}). These results indicate that even if the density fluctuations can noticeably change at close to the Sun distances they do not seem to significantly vary in the inner heliosphere.

It is generally considered that the density fluctuations can influence the level of the scattering of the radio emission, which then subsequently also affects the accuracy of the estimated radio source positions and the sizes of the radio sources \citep{Kontar19, Krupar20}. In light of our results on the density fluctuations, obtained from the in situ observation, we can consider that the scattering effects do not significantly change at the distances considered in this study. This conclusion is also in accordance with our findings (see Fig.~\ref{Fig: distances}),which shows no systematic variation of the radio source sizes obtained in this study, contrary to the results obtained in the metric wavelength range \citep{Saint-Hilaire13}.

\begin{figure*}
    \centering
    \includegraphics[width=18cm]{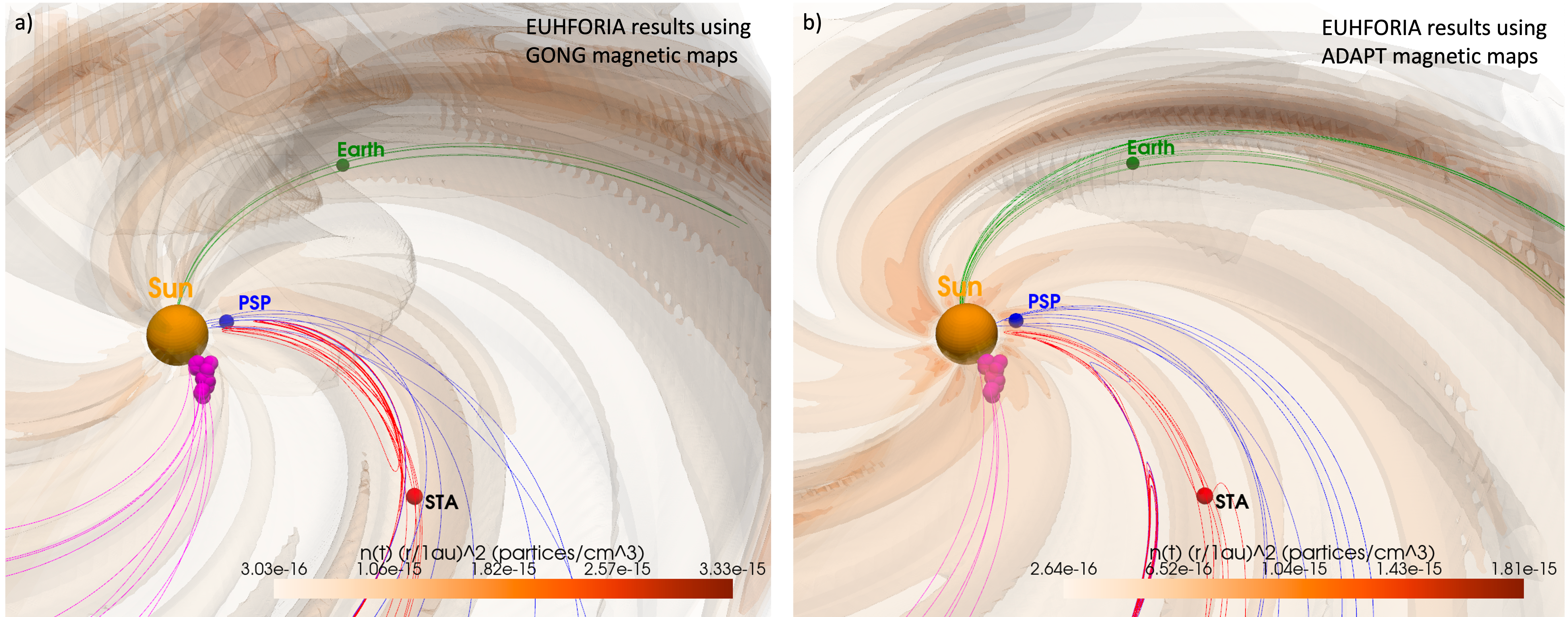}
     \caption{3D Iso-surface plot of the scaled density obtained from EUHFORIA simulations. Sphere at the center represent the inner coronal boundary of EUHFORIA at 0.11 AU. Magenta spheres represent the radio source positions of the Event 8. The radio sources are not plotted in their actual sizes. Blue, red and green spheres represent PSP, STEREO-A and Earth positions respectively. The magenta, red, blue, and green lines represent the magnetic field lines connecting the radio source at 775 kHz, STEREO-A, PSP, and Earth to the Sun, respectively. Panel (a) and (b) shows 3D representation of the EUHFORIA density obtained using GONG and ADAPT magnetograms respectively.}
              \label{Fig: EUHFORIA3D}
    \end{figure*}

\subsection{Uncertainties and inaccuracies of solar wind density through modeling and observations}
\label{sec:uncertainities}

Both, modeling and observations can be influenced by a spectrum of different factors that can in different ways introduce uncertainties and inaccuracies into the results. Knowledge on these factors is crucial for improving the interpretation of the results and their significance, derived from both modeling and observational studies. We will here address some of the factors that influence the accuracy of results of our study, grouped as uncertainty arising from: a) modeling, b) radio and in situ observations (already partially discussed in the Section~\ref{sec:fluctuation}).

\subsubsection{Modeling uncertainties}
\label{sec:model uncertainities}

We will first address the uncertainty in the estimation of the EUHFORIA density. In general uncertainties in modeling can arise due to different factors and we describe only the most important one such as the model set up, the limitations in the input data, modeling assumptions and modeling limitations in a case of complex events.

The main input to EUHFORIA are $360^{\circ}$ magnetic synoptic maps which are available from different providers and can differ significantly.
As a result, such diverse synoptic maps can strongly influence modeled solar wind characteristics \citep{Riley12, Riley14, Riley21, Perri23}. 
Further, synoptic magnetograms are constructed using the `old' information for the far-side of the Sun. This is particularly important when modeling solar wind and the events at/behind the East solar limb, for which the magnetogram input is almost half solar rotation old. We note that this was exactly the case in our study. These uncertainties could be fully addressed only if the $360^{\circ}$ observations of the Sun are available.

The EUHFORIA results also showed that the use of different magnetic maps, namely GONG and ADAPT, can lead to different simulation results (Fig.~\ref{Fig: EUHFORIA_Time_series}a \& b). Figure~\ref{Fig: EUHFORIA2D} shows the equatorial and meridional slices of the EUHFORIA domain obtained using GONG and ADAPT synoptic maps, respectively. Although both results display high density regions along the PSP propagation path (fig.~\ref{Fig: EUHFORIA2D}a~\&~c), the overall density structures differ from each other.

Another factor that gives rise to the uncertainty in the modeling results is the simplicity of the EUHFORIA's coronal model \citep{Hinterr19}. Some of the recent studies indicate that more advanced coronal models, than the herein used default coronal model of EUHFORIA, may increase the accuracy of EUHFORIA \citep{Samara21}. Additionally, the more accurate background solar wind is expected in the case of time-dependent modeling \citep{Baratashvili2025}. 

Furthermore, EUHFORIA strongly relies on the solar wind boundary conditions used as an input to the heliospheric model, such as fast solar wind density and velocity \citep{Pomoell18}. The improvements of this modeling input could be still expected from the upcoming in situ solar wind observations by the current novel missions such as PSP and SolO \citep[Solar Orbiter;][]{Muller20}.

\subsubsection{Uncertainties induced with in situ \& radio observations}
\label{sec:radio uncertainities}

Since 2018 and its first close approach PSP provides us with a large amount of close-to-the-Sun in situ observations. As PSP orbits the Sun and approaches its perihelion, it maps the local coronal plasma properties in high temporal resolution at constantly changing positions. PSP in situ observations show that, in general, the electron density of the coronal plasma vary by about one order of magnitude \citep{Kruparova23}. Therefore, when comparing the electron density obtained from radio observations with those from PSP in situ measurements, it is important to understand that different position and trajectories of PSP and radio sources will probably result in different coronal electron densities. For more detailed discussion see also Section~\ref{sec:fluctuation}.

The dynamic spectra from PSP provide valuable information regarding observed type III radio bursts. With PSP's high-resolution data we can identify multiple type III bursts that are not fully resolved in the dynamic spectra from Wind or STEREO due to their lower temporal resolution. Sometimes, type III bursts occurring subsequently within a short time interval of e.g. few minutes can be identified in Wind and STEREO dynamic spectra as a group of bursts which are not fully resolved. On the other hand the time series of the same type III radio bursts at discrete frequency will show only complex time profile, suggesting the presence of more than one type III radio bursts \citep{Gerekos24} but without any certainty. This difference arises from the additional information provided with large frequency range in dynamic spectra. The complex time profile is also accompanied with the fluctuating values of azimuth and colatitude of the DF observations, leading at the end to decreased accuracy in estimation of the radio source positions (Deshpande et al., in prep.). In order to avoid this uncertainty we selected type III burst as isolated as possible. Additionally, when selecting the time series points for triangulation we aimed at those that had the least fluctuations of the azimuth and colatitude components for the subsequent measurements and were close to the peak intensity of the burst time profile. We need to note that in the case of our study the direction of propagation of the radio sources obtained from the DF generally stays conserved (Figs.~\ref{Fig: Radio Position} and \ref{Fig: all_typeIII}) as the scattering effects do not seem to be significant. The confirmation of that, beyond here presented study and other already published works \citep{Magdalenic14, Jebaraj20} which shows the similar results, is also visible in the Fig.~\ref{Fig: EUHFORIA3D} of the manuscript where it is clearly shown that the type III radio sources propagate along the magnetic field lines obtained independently using the 3D MHD EUHFORIA code.

PSP measurements reveal sometimes very large fluctuations of the local electron density highlighting the turbulent nature of the solar corona. Given this highly dynamic behavior of the solar wind plasma, accurately estimated radio source positions are crucial for mapping correct electron densities. Accurate estimation of the radio source sizes is also important, as errors in source size estimation can result in incorrect density assumptions for the large solar wind regions. To decrease the level of these uncertainties we need to try to avoid the bursts with the complex time profiles.

\section{Discussion and conclusions}
\label{sec:conclusion}
 
This study presents the radio triangulation of interplanetary type III bursts and characteristic of the obtained radio sources.
The results of analysis of radio bursts is employed in the study focused to the coronal electron density obtained using different methods. We inspect electron density estimated with type III radio bursts, novel PSP observations and EUHFORIA modeling results. The main focus of the study is to map the electron density in the interplanetary space, as accurate as possible, and inter-validate the results obtained with radio observations and EUHFORIA.

Our study encompasses 11 type III radio bursts observed during the second PSP perihelion. The selected bursts were required to be intense, isolated from the other types of bursts and observed by both Wind and STEREO-A spacecrafts. STREREO-A and PSP instruments observed multiple type III radio bursts during the 2nd PSP perihelion. However, only few of them were observed by Wind. This might be because of the different propagation path of the bursts and/or the low temporal resolution of the Wind instrument. This condition restricted the selection of the bursts for which triangulation can be applied.

The studied radio events were associated with the AR (NOAA AR 12738) located behind the east solar limb. The EUV observations show association between EUV jets observed in all four frequency channels of STEREO-A/EUVI and accelerated electrons which produced radio bursts. Such a scenario was also reported in some previous studies \citep[e.g.~][]{Bain09, Klassen11, Krucker11, Mccauley17, Cairns18, Mulay19}. The recent work which focused on type III radio bursts observed within the same period around second PSP perihelion \citep{Pulupa20, Krupar20, Harra21, Stanislavsky22, Wang23, Nadal23}, showed that the type III bursts were associated with the jets from the AR 12738 which also agrees with our findings herein. 

Understanding the characteristics of the time profiles is crucial for assessing the reliability of radio triangulation results. Namely, presence of the close in time subsequent burst, or the simultaneous presence of the fundamental and harmonic component will induce the inaccuracy in the triangulation results. Therefore, in order to understand how precise our results are, we analyzed the time profiles of selected bursts observed by STEREO-A and Wind (Fig.~\ref{Fig: Time Profiles}). STEREO-A showed complexity in the time profiles confirming the presence of more than one type III radio bursts in a few studied events (Event 7, 10 and 12). On the other hand Wind observations of the same bursts showed a single-peak time profile. This can be also due to the low intensity and/or low temporal resolution of Wind data.

Results for the Event 8 and Event 9 are presented in detail in this paper. Radio triangulation method provided us 3D radio source positions and allowed us to map the direction of propagation of the type III bursts which was not along the direction of the Parker spirals. Event 8 showed downward propagation path from the ecliptic plane, whereas Event 9 showed direction of propagation along the ecliptic plane (Fig.~\ref{Fig: Radio Position}). This result shows that the propagation path of the subsequent type III radio bursts observed within few minutes and originating from the same AR was along different magnetic field lines (Magdalenic et al., in prep.). We, note that even if the subsequent type III bursts are originating from the same AR they might not have the same propagation path. \cite{Jebaraj20} showed that two groups of type III bursts from the same radio event propagated along the different flanks of the CME and very distinct magnetic field lines. Herein presented results and the one by \cite{Jebaraj20} show that contrary to the generally considered fact, propagation path of subsequent type III bursts can be very different, following distinctively different field lines and not following Parker spiral. On the other hand, all studied type III radio bursts which originate from the same AR in the course of 4 days have comparable propagation path as visible from Fig.~\ref{Fig: all_typeIII}a. Moreover, subsequent bursts seem to follow the propagation of the AR, with the apparent movement of the AR across the solar disk as observed from STEREO-A, showing the angular spread when viewed in the ecliptic plane (Fig.~\ref{Fig: all_typeIII}b). 

We found that radio source size, defined as the shortest distance between the wave vectors (Fig.~\ref{Fig: illustration}), ranges from \mbox{0.5\,--\,40}~$\deg$ or \mbox{0.5\,--\,25}~R$_\odot$ (Fig.~\ref{Fig: distances}). Obtained source sizes do not exhibit any clear dependence on the observing frequency, suggesting that scattering effects are not significant in this frequency range and in the case of studied events. This finding indicates that while radio wave scattering can influence observed source positions, its impact might be less significant at the distances and frequencies considered in this study. The radio source positions at the higher frequencies cluster closer to the Sun while the low frequency sources (e.g. 425~kHz) appear further away from the Sun, as see in Fig.~\ref{Fig: all_typeIII}b. This behavior follows from the direct proportionality of the plasma frequency and coronal electron density (Eq.~\ref{eq:2}). 
We note that even if study is based on the 11 type III bursts which is not a large number at each triangulation frequency pair several data points were considered resulting in the significant number of radio source positions and sizes (Fig.~\ref{Fig: distances}) making our results also statistically firm.

The radio triangulation provided 3D radio source positions which were used to map the electron density in the inner heliosphere (\autoref{eq:2}). The radio density when compared with the 1D models (\citealp{Newkirk61, Saito70, Leblanc95, Vrsnak04, Kruparova23}; Fig.~\ref{Fig: Density model}), were found to be along the 20~$\times$~Leblanc model with the lower edge comparable to 3.5~$\times$~Saito. The average PSP densities over the 10 first perihelion passages \citep{Kruparova23} are about  an order magnitude lower than the radio density profiles. One of the reasons for the obtained difference is probably different speed and propagation path of the type III bursts and PSP. The type III radio bursts propagate with the speed of a fraction of speed-of-light along the magnetic filed lines, approximately radially away from the Sun. In the same time PSP propagates significantly slower than radio bursts and more perpendicular to the solar wind flow. Therefore, PSP and radio bursts encompass very different portions of the corona during the type III duration. While single type III bursts encompasses distances from about 0.12~au down to 0.2~au, PSP encompasses only about 0.0008~au. Therefore, the variation of PSP density is expected to be low in comparison to the radio density. 

Another reason for discrepancy between radio and PSP density is still notably different starting position of PSP and type III radio bursts. The combined analysis of the SOHO/LASCO and STEREO COR2 images (Fig.~\ref{Fig: Coronagraph}) with EUHFORIA modeling results (Fig.~\ref{Fig: EUHFORIA2D} panels (a) \& (b), and Fig.~\ref{Fig: EUHFORIA3D}) show strong variations and structuring of the solar wind velocity/density in the regions of PSP and type III radio sources. At the beginning of the perihelion passage PSP propagates across the regions of slow solar wind probably originating from the equatorial streamer belt region composed of at least 3 distinguishable streamers stretching outwards in the radial direction (Fig.~\ref{Fig: Coronagraph}). Starting from the 03~April and during the Event 8, PSP propagates in the fast solar wind characterized with the speed reaching up to 600 km/s and with the low density of about few hundred particles/cm$^3$, as seen from Fig.~\ref{Fig: EUHFORIA_Time_series}. The solar wind density fluctuations mapped by PSP show variation of very local plasma condition in PSP vicinity. On the other hand the type III radio sources which propagate along the magnetic filed lines, seem to be situated within different slow solar wind stream which also indicates different density regime. We concluded that strong structuring of the solar wind density which is mapped very locally by both radio sources and PSP, also attributes to the significant difference between the radio density and the PSP densities for the studied events. 

The space weather model EUHFORIA was employed to estimate the densities at PSP and radio source positions for Event 8 (as sketched in Fig.~\ref{Fig: position}b). Initially, we used GONG magnetograms as input to EUHFORIA to estimate the densities at PSP position. The simulations performed with the default PFSS SSR of 2.6~$R_{\odot}$ underestimated the densities compared to PSP in situ observations (Fig.~\ref{Fig: EUHFORIA_Time_series}a). To exclude the possibility of opening the magnetic field lines that are in reality closed and considered as such when using the default PFSS SSR, we gradually increased the SSR to 3.0~$R_{\odot}$. Increasing SSR resulted in a better agreement with the in situ densities at PSP position but still did not fully match PSP density. Therefore, we employed ADAPT magnetic maps as input to EUHFORIA using the default PFSS SSR value. All ADAPT realizations yielded higher densities compared to those obtained with GONG magnetic maps, and they showed better agreement with PSP in situ density (Fig.~\ref{Fig: EUHFORIA_Time_series}b). 

We then compared the radio densities at all direction-finding frequencies with the EUHFORIA densities at the corresponding radio source positions for Event 8. The radio densities were higher by one order of magnitude than PSP in situ densities. Whereas, the densities obtained with EUHFORIA at the radio source positions are observed to be lower than the one obtained with the radio and in situ observations (Fig.~\ref{Fig: EUHFORIA_Time_series} c to h). Despite the obtained discrepancies EUHFORIA did predicted denser region along the propagation path of the type III bursts (Fig.~\ref{Fig: EUHFORIA2D}). 

The comparison of in situ, radio, and EUHFORIA density revealed differences which pointed to the necessity to discuss uncertainties associated with employed methods of density estimation. The discrepancy in the results could be attributed to several factors. One key aspect is the modeling process itself, as EUHFORIA relies heavily on the provided input data.  The most important parameter is the synoptic magnetic field map used as the main model input. It is needed to note that all the events analyzed in this study had source region located behind the east solar limb. Accordingly, the magnetic field information was almost half solar rotation old. This uncertainty can be only corrected by employing $360^{\circ}$ magnetic field observations of the Sun.

The accuracy of EUHFORIA’s modeling results is also influenced by the simplicity of its coronal model, with recent studies suggesting that more advanced models should be used \citep{Hinterr19, Samara21}. Additionally, time-dependent modeling is expected to further improve the accuracy of the background solar wind predictions. EUHFORIA's reliance on solar wind boundary input parameters, such as fast wind density and velocity, means that novel in situ observations from missions like PSP and SolO could provide more realistic boundary conditions for EUHFORIA. 

PSP’s high-resolution in situ measurements have shown that coronal electron densities vary by an order of magnitude, emphasizing the importance of considering different positions and trajectories when comparing radio and in situ density measurements. Furthermore, PSP's dynamic spectra indicated the presence of multiple type III bursts that generally remain undetected in the lower-resolution data from Wind and STEREO. This findings emphasis the complexity of the studied signals and potential inaccuracies in radio source positioning through triangulation. Given the highly dynamic and turbulent nature of the solar corona, precise identification of mapped regions is crucial for density measurements. Moreover, errors in estimating radio source sizes can lead to incorrect density assumptions, introducing additional uncertainties in the results.

Our analysis of PSP electron density variations indicates that small-scale RMS density fluctuations are around 0.05. When considering a longer time window of 10 minutes, which encompasses the full duration of a single type III radio burst, the RMS density fluctuations increase to approximately 0.06. These results suggest that while density fluctuations can impact radio wave scattering and therefore also radio source positions, they do not show significant variation across the inner heliosphere at the distances analyzed in this study. 

We note that both independently obtained results; a) on the behavior of the radio source sizes obtained in radio triangulation and the b) on the density fluctuations from the PSP observations indicate low level of scattering effects for the studied events. Similar results on the scattering effects for the metric range radio bursts were recently obtained by \citet{Kumari25}.

The main conclusions of the performed studies are listed here:
\begin{itemize}[label=\textbullet]

   \item Radio triangulation results of Event 8 showed propagation path of the radio sources southward from the ecliptic plane, whereas, radio sources in Event 9 propagated along the ecliptic plane. We conclude that the subsequent type III bursts can propagate along different propagation path and along different magnetic filed lines, even though they are generated from the same AR. 
   \item When all the studied events were plotted together, the majority showed a southward propagation path from the solar ecliptic plane. This suggests that presence of different long-living coronal structures, such as coronal streamers, and dynamic changes in the ambient coronal magnetic field configuration caused by solar activity, such as CMEs, can significantly influence the propagation path of type III radio bursts.
   \item We have found that the radio source sizes were relatively small compared to some previous studies and in the range of \mbox{0.5\,--\,40}~$\deg$ or between \mbox{0.5\,--\,25}~R$_\odot$. Radio source sizes did not show systematic variation with the frequency and therefore, scattering which is expected to directly influence radio source positions and sizes, as a function of the local electron density, might not be significant in this frequency range. This conclusion is confirmed with the low level of the density fluctuations, generally responsible for the scattering effects, obtained from the PSP in situ data.
   \item The uncertainties on the coronal electron densities obtained from comparison of radio, PSP and EUHFORIA densities are mainly due to :
   \begin{itemize}
       \item Different propagation path of type III radio bursts and PSP indicates that we are mapping different coronal regions. 
       \item Type III radio bursts might be propagating along coronal streamer lines, which then explains the high radio densities.
       \item AR associated with type III radio bursts was behind east solar limb. Consequently, magnetic maps were constructed with almost half solar rotation old information which strongly influences the accuracy of the modeling results.
       \item Magnetograms obtained from different providers can induce very different modeling results.
       \item Simplicity of the coronal model of our MHD code EUHFORIA \citep{Pomoell18} and time independent nature of the model might contribute the model accuracy.
   \end{itemize}
   \item Despite the relative differences in the three types of electron densities, we can note that, EUHFORIA rather successfully identified regions of increased density along the propagation paths of type III radio bursts, demonstrating its capability to capture broader density structures in the heliosphere.
   \item  The obtained behavior of the radio sources and the density fluctuation analysis of PSP in situ data showed that the scattering effects do not change significantly at the distances considered in this study. 
\end{itemize} 

The current high level of the solar activity is expected to provide us with PSP and SolO observations which will bring us new opportunities to expand this work even further. Our study was limited to the second perihelion due to restrictions in the separation angle of STEREO-A and Wind, which are crucial for radio triangulation. However, future observations will benefit from more favorable spacecraft alignments, allowing for the analysis of a larger number of events. Additionally, PSP’s improved time resolution of 0.3 seconds will facilitate a more detailed investigation of radio events. The newly launched PROBA-3 mission \, along with ground-based observational opportunities, will further enhance our ability to probe the lower corona, providing deeper insights into the origin and propagation of solar radio emission. The recently developed coronal model COCONUT \citep{Perri22, Perri23}, will, together with the time-dependent version of EUHFORIA provide more accurate modeling results. We believe that these advancements will even more improve our understanding of electron density variations and the physics of the solar radio bursts in the inner heliosphere.

\begin{acknowledgements}

K.D. is supported by the PhD grant awarded by Royal Observatory of Belgium.
The authors also acknowledge financial support by the BRAIN.be project SWIM, University start-up grant 3E220031 and the FEDtWIN project PERIHELION.
The ROB team thanks the Belgian Federal Science Policy Office (BELSPO) for the provision of financial support in the framework of the PRODEX Programme of the European Space Agency (ESA) under contract numbers 4000112292, 4000134088, 4000106864, 4000134474, and 4000136424.
J.M. and I.C.J. were supported by the International Space Science Institute (ISSI) in Bern, through ISSI International Team project No.557, “Beam-Plasma Interaction in the Solar Wind and the Generation of Type III Radio Bursts”.
I.C.J. is grateful for support by the Research Council of Finland (X-Scale, grant No.~371569). I.C.J also acknowledges funding from the European Union's Horizon Europe research and innovation programme under grant agreement No.\ 101134999 (SOLER). The paper reflects only the authors' view and the European Commission is not responsible for any use that may be made of the information it contains.
\\
Authors would like to thank all the teams responsible for the space based observations. The data are publicly available at \url{https://spdf.gsfc.nasa.gov/} (PSP, STEREO, Wind).
We would like to thank JHelioviewer for being able to browse LASCO, SECCHI, AIA and EUVI data  \citep[\url{http://jhelioviewer.org};][]{Muller17}. We also thank provider of magnetograms, i.e., NSO/GONG and ADAPT maps (\url{https://gong.nso.edu/data/magmap/}) and \url{https://nso.edu/data/nisp-data/adapt-maps/}, respectively). \\
The authors thank the anonymous reviewers for their helpful comments that enhanced the value of this paper.

\end{acknowledgements}

\bibliographystyle{aa} 
\bibliography{bibtexKetaki}

\end{document}